\documentclass[lettersize,journal]{IEEEtran}
\usepackage{amsmath,amsfonts}
\usepackage{algorithmic}
\usepackage{algorithm}
\usepackage{array}
\usepackage[caption=false,font=footnotesize,labelfont=rm,textfont=rm]{subfig}
\usepackage{textcomp}
\usepackage{stfloats}
\usepackage{url}
\usepackage{verbatim}
\usepackage{graphicx}
\usepackage{cite}
\usepackage{amsmath}
\usepackage{amssymb}
\usepackage{bm}
\usepackage{color}
\allowdisplaybreaks[4] 
\begin{document}

\title{Antenna Orientation Optimization for\\ Rotatable Antenna-Enabled ISAC Systems}

\author{Qingjie Wu, Beixiong Zheng,~\IEEEmembership{Senior Member,~IEEE}, Guangchi Zhang, and Robert Schober,~\IEEEmembership{Fellow,~IEEE} \vspace{-0.1cm}
	\thanks{		
		\textit{(Corresponding author: Beixiong Zheng.)}
		
		Qingjie Wu and Beixiong Zheng are with the School of Microelectronics, South China University of Technology, Guangzhou 511442, China (e-mail: miqjwu@mail.scut.edu.cn; bxzheng@scut.edu.cn).
		
		Guangchi Zhang is with the School of Information Engineering, Guangdong University of Technology, Guangzhou 510006, China (e-mail: gczhang@gdut.edu.cn).
		
		Robert Schober is with the Institute for Digital Communications, Friedrich-Alexander-University Erlangen-N$\ddot{\mathrm{u}}$rnberg (FAU), 91054 Erlangen, Germany (e-mail: robert.schober@fau.de).
		}
}

\maketitle

\begin{abstract}
Rotatable antenna (RA) has emerged as a promising technology to improve both communication and sensing performance in future wireless networks. In this paper, we deploy an RA array at the base station (BS) to improve the integrated sensing and communication (ISAC) performance by exploiting the additional spatial degrees of freedom (DoFs) introduced by antenna rotation.
To enhance the sensing performance over an extended region containing a potential target while meeting the communication requirements of multiple users, we aim to maximize the minimum echo signal power within the sensing region, subject to required minimum communication rates of the users.
For the special case of a single user and a point target, we develop a tractable approximation of the original problem, leading to independent per-RA subproblems. We characterize the geometric structure of these subproblems and derive closed-form solutions. Under the far-field approximation, the per-RA subproblems become approximately identical, allowing a common pointing vector to be adopted for all RAs.
For the general multi-user and regional sensing case, we propose an alternating optimization (AO) algorithm that alternately optimizes the transmit beamforming for communication, the covariance matrix of the probing signal, and the pointing vectors of the RAs in an iterative manner.
Simulation results demonstrate that the proposed RA-enabled ISAC system significantly outperforms various benchmark schemes with restricted antenna-orientation flexibility, while the evaluation of the Cramér-Rao bound (CRB) further verifies its benefits for target-position estimation accuracy.
\end{abstract}

\IEEEpeerreviewmaketitle

\section{Introduction}\label{Introduction}

\subsection{Background}
Integrated sensing and communication (ISAC) is widely regarded as one of the most important application scenarios and enabling technologies for the sixth-generation (6G) wireless network, as it unifies the communication and sensing functionalities by sharing the same hardware and spectrum resources~\cite{Liu2022Integrated,Chowdhury20206G}.
This integration reduces deployment costs and energy consumption, enabling a broad range of applications in location-aware scenarios, such as the low-altitude economy, immersive virtual reality, and vehicle-to-everything (V2X) communications~\cite{Saad2020A}.
To facilitate ISAC and satisfy its diverse requirements, multiple-input multiple-output (MIMO) is recognized as a key enabling technology, owing to its capabilities in spatial adaptation and waveform design through multiplexing and beamforming~\cite{Liu2022A,Hua2024MIMO}, while emerging receiver technologies, such as Rydberg atomic receivers, are also being explored as promising sensing front ends for future wireless systems~\cite{Chen2026Harnessing,Chen2026Polarization}.
However, existing MIMO-enabled ISAC systems mainly rely on traditional fixed-antenna architectures, which cannot adaptively respond to the variations of wireless channels and user distributions. As a result, they may limit the achievable performance and overall potential of ISAC systems, especially in highly dynamic environments.
Additionally, to meet the growing demand for both communication and sensing, future ISAC systems are expected to evolve towards large-scale antenna arrays. However, this trend inevitably leads to increased hardware costs, which poses a significant challenge to the development of cost-effective and high-performance ISAC systems.

\vspace{-0.02cm}
To overcome the limitations of conventional fixed-antenna ISAC systems, several recently proposed reconfigurable antenna architectures, such as fluid antenna system (FAS)~\cite{New2024A}, movable antenna (MA)~\cite{Zhu2024Movable}, six-dimensional MA (6DMA)~\cite{Shao2025Tutorial}, and rotatable antenna (RA)~\cite{Zheng2025RotatableMAG}, can be leveraged to enhance the performance limits of ISAC.
Specifically, since FAS and MA can proactively exploit spatial diversity and favorably reshape wireless channels by flexibly adjusting antenna positions, they provide enhanced spatial multiplexing capabilities for communication and high angular resolution for sensing~\cite{Zhou2024Fluid,Wang2024Fluid,Chen2025Antenna,Ding2026Movable}.
Meanwhile, the real-time adjustment of antenna positions enables ISAC systems to adapt to time-varying environmental conditions and heterogeneous communication and sensing requirements~\cite{Ma2024Movable,Kuang2024Movable}.
Therefore, FAS and MA are promising technologies for ISAC applications, as they offer increased design flexibility and enable performance gains for both functionalities. Nonetheless, the ISAC performance gain achieved with MA is limited as MAs cannot perform array rotation.
To address this issue, 6DMA has emerged as a potential enabling technology for fully exploiting the spatial degrees of freedom (DoFs) of ISAC systems by flexibly adjusting both the three-dimensional (3D) positions and 3D rotations of distributed antenna arrays~\cite{Shao2025Exploitin,Wang2025UAV}.
Besides enhancing the array gain for communication users, 6DMA can also improve sensing and localization accuracy by exploiting the geometric gain inherently offered by distributed 6DMA surfaces, as the relative geometry between the target and the 6DMA-based base station (BS) can be adaptively configured~\cite{Shao20256DMA}.
Therefore, by properly deploying the 6DMAs based on the communication and sensing requirements as well as the spatial distributions of the users and targets, the overall ISAC performance can be significantly enhanced. Nevertheless, the 6DMA architecture may incur high hardware overhead and system design complexity. These drawbacks thus call for the development of a flexible antenna architecture that can enable cost-effective ISAC while preserving sufficient spatial adaptation capability.

\subsection{RA-Enabled ISAC}
RA technology~\cite{Wu2024Modeling,Zheng2025Rotatable,Zheng2026Tutorial} has recently emerged as a new antenna architecture that focuses exclusively on antenna orientation/boresight rotation flexibility while keeping the antenna position fixed. It offers a cost-effective and compact solution and can be regarded as a simplified yet promising member of the MA/6DMA/FAS family.
By independently rotating the 3D orientation/boresight of each RA, the array directional gain pattern can be reconfigured to meet the performance requirements of various application scenarios, such as secure communication~\cite{Dai2025Rotatable}, spectrum sharing~\cite{Peng2026Rotatable,Tan2026Rotatable}, and cell-free communication~\cite{Pan2025Rotatable,Peng2026Cell}.
From a communication perspective, the RA architecture enables dynamic rotation of the antenna mainlobes according to the spatial distribution of the users, thereby providing more refined spatial beam management. Specifically, it can enhance the array gain in desired directions while reducing radiation leakage in undesired directions, which helps strengthen the links of intended users, mitigate inter-user interference, and improve channel capacity as well as communication reliability in multi-user systems~\cite{Wu2024Modeling,Zheng2025Rotatable,Zheng2026Tutorial}.
From a sensing perspective, RA offers flexible directional control to accommodate different sensing objectives. When the orientations/boresights of multiple RAs are aligned toward the same target or region, the directional gains of the RAs can be leveraged for improving target illumination and enhancing the received echo signal power. In contrast, when the RA orientations/boresights point in different directions, the sensing region can be effectively enlarged, thereby enabling wider sensing coverage. Therefore, the rotational DoFs of an RA array provide additional flexibility to balance directional sensing gain and spatial sensing coverage given the sensing tasks at hand.

As such, the rotational DoFs offered by RA open up new opportunities to achieve significant gains over traditional fixed-antenna architectures in ISAC systems, by providing enhanced spatial adaptability and improved signal directivity without the need for increasing the number of antennas~\cite{Wu2025Rotatable,Xiong2026Intelligent,Zhou2025Rotatable,Zhang2026Rotatable,Qu2025Rotatable}.
Therefore, by jointly optimizing the antenna orientations/boresights together with the communication beamforming and probing signals, the RA architecture is expected to achieve a more flexible trade-off between communication and sensing performance while simultaneously improving both spectral and energy efficiency.

\subsection{Motivations and Contributions}
Although RA-enabled ISAC has been investigated in several existing studies, such as \cite{Zhou2025Rotatable,Zhang2026Rotatable,Qu2025Rotatable}, these works mainly focus on array-wise rotation, which provides only coarse-grained orientation control for balancing the array aperture gain between communication users and sensing targets. However, array-wise rotation may become inadequate in more complex scenarios, especially when multiple communication users need to be served simultaneously or when sensing targets are spatially dispersed,
since all antennas share a common orientation and thus cannot independently adapt their directional gains to heterogeneous spatial demands. In practical sensing scenarios, the exact target location may not be known a priori but only confined to a region of interest, which calls for maintaining satisfactory sensing performance over an entire region rather than concentrating the sensing gain on a prescribed location. In this context, compared with array-wise rotation, element-wise RA rotation provides fine-grained spatial controllability by independently adjusting the orientation of each RA, thereby introducing additional spatial DoFs to flexibly shape the effective radiation distribution according to the spatial distribution of multiple communication users and the size and shape of the sensing region. However, the potential of element-wise RA systems for multi-user communication and regional sensing has remained largely unexplored. This motivates this work.

In this paper, we investigate a new RA-enabled ISAC system, as shown in Fig.~\ref{fig_system}, in which an element-wise rotation mechanism is introduced to provide additional spatial DoFs for improving system performance. Specifically, an RA array is deployed at a dual-functional BS to simultaneously communicate with multiple users in the downlink and sense a potential target randomly located within an extended region by receiving an echo signal in the uplink.
To achieve the ISAC functionality and enhance regional sensing capability, the transmit ISAC beam and the orientation of each RA are jointly designed to improve sensing performance over the entire region where the target may be located, while guaranteeing the communication performance of each user.
Therefore, we aim to maximize the minimum echo signal power of the potential target within the sensing region by jointly optimizing the transmit beamforming for communication, the probing-signal covariance matrix, and the orientations of all RAs, while ensuring that the communication rate of each user does not fall below a given threshold.
The main contributions of this paper are summarized as follows:
\begin{figure}[!t]  \centering
	\includegraphics[width=2.2in]{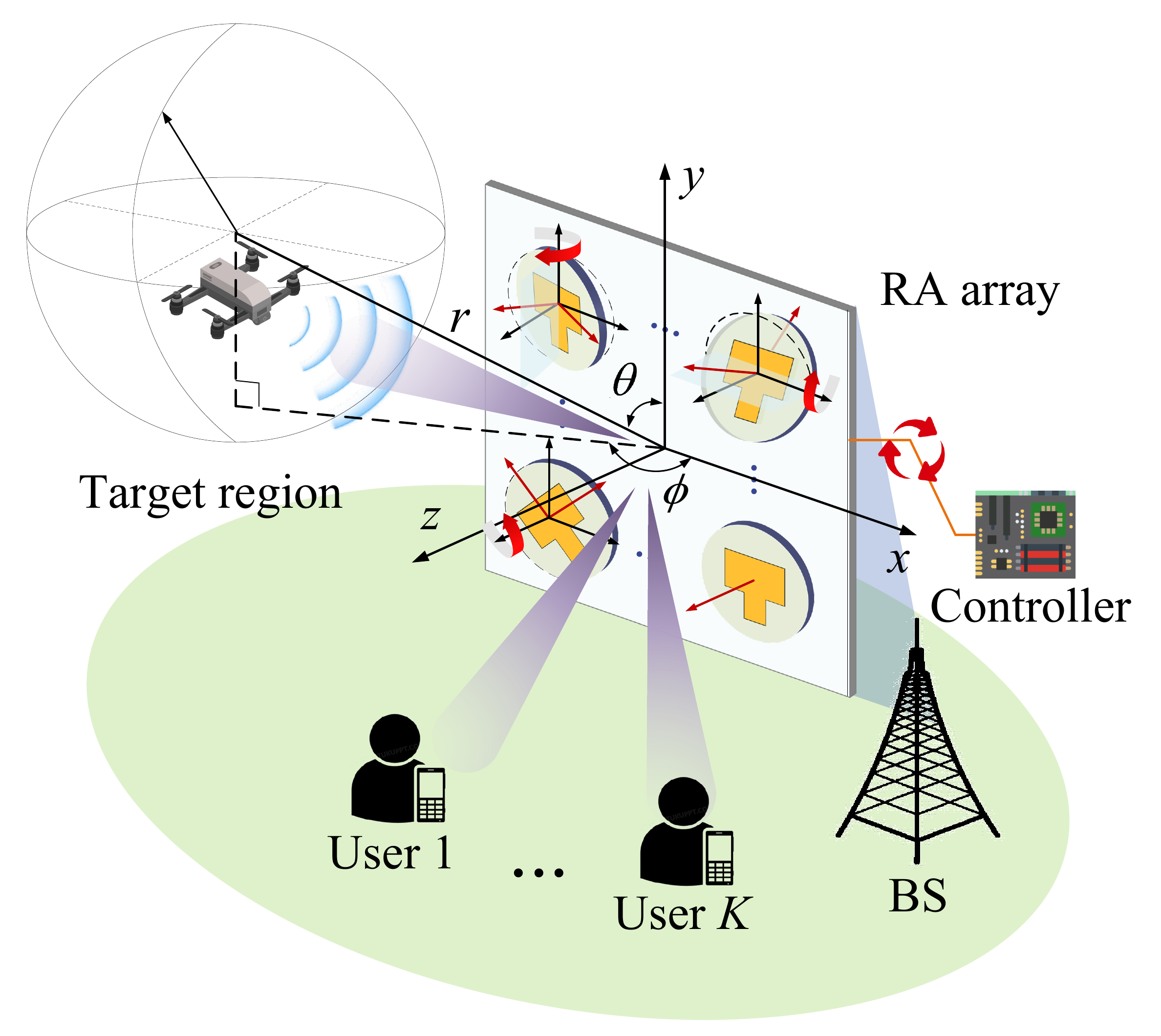}\vspace{-0.2cm}
	\caption{An RA-enabled ISAC system for multi-user communication and regional sensing.}
	\label{fig_system}\vspace{-0.3cm}
\end{figure}
\begin{itemize}
	\item{We first model the impact of the antenna orientation/boresight rotation on the communication channel and the round-trip sensing channel. Different from communication-oriented RA designs and array-wise RA-enabled ISAC designs, the independently adjustable RA pointing vectors can simultaneously accommodate multiple user directions and an extended sensing region. Based on the resulting channel models, we formulate a max-min regional sensing problem to maximize the worst-case received echo signal power across the entire sensing region, subject to minimum communication rate requirements of all users.}
	\item{For the special case of a single user and a point target, we develop a tractable approximation for the RA pointing design, which can be decomposed into independent per-RA subproblems, whose geometric structure we analytically characterize. We show that the optimal pointing vector of each resulting per-RA subproblem lies either within the sector spanned by the user and target directions or on the boundary of the allowable rotation cone, and derive closed-form optimal solutions for both cases. The resulting expressions reveal how the communication requirement, the user-target geometry, and the allowable rotation range jointly determine the RA pointing direction.}
	\item{For the general multi-user and regional sensing case, the continuous sensing region is discretized into spatial sampling points, thereby converting the semi-infinite regional sensing requirement into a finite set of constraints. To efficiently solve the resulting non-convex joint design problem, we develop an alternating optimization (AO)-based solution framework, where standard semidefinite relaxation (SDR) and successive convex approximation (SCA) techniques are employed to optimize the ISAC signals and the element-wise RA pointing vectors, respectively.}
	\item{Simulation results are provided to evaluate the proposed optimization algorithm, demonstrating that the RA architecture can achieve much better ISAC performance than various benchmark schemes. The results show that the proposed RA-enabled ISAC system can effectively enlarge the region characterizing the trade-off between communication and sensing, and achieve a lower Cramér-Rao bound (CRB) for target estimation, validating its superiority from the perspective of estimation accuracy. Moreover, the results for extended sensing region reveal an important trade-off between directional sensing gain and sensing coverage: higher antenna directivity is beneficial for compact sensing regions, whereas excessive directivity may degrade the worst-case sensing performance over large regions due to insufficient beam coverage.}
\end{itemize}

The remainder of this paper is organized as follows. Section~\ref{System_model} presents the system model and problem formulation for the considered RA-enabled ISAC system. In Section~\ref{Single_user}, we characterize the geometric structure of the RA pointing design and derive closed-form solutions for the resulting per-RA subproblems for the single-user and point-target case. In Section~\ref{Multi_user}, we propose an AO algorithm to solve the formulated problem in the multi-user and regional sensing scenario. Simulation results are presented in Section~\ref{Simulation} to evaluate the performance of the proposed design and algorithm. Finally, Section~\ref{Conclusion} concludes the paper.

\textit{Notation:} Upper-case and lower-case boldface letters denote matrices and column vectors, respectively. Superscripts $(\cdot)^T$, $(\cdot)^{\ast}$, $(\cdot)^H$, and $(\cdot)^{-1}$ stand for the transpose, conjugate, Hermitian transpose, and matrix inversion operations, respectively. The sets of $a\times b$ dimensional complex and real matrices are denoted by $\mathbb{C}^{a\times b}$ and $\mathbb{R}^{a\times b}$, respectively. $|\cdot|$ denotes the absolute  value, $\odot$ denotes the Hadamard product, and $\mathbb{E}\{\cdot \}$ denotes the expectation of a random variable. $\mathcal{O}(\cdot)$ is the standard big-O notation. For a vector $\bm{x}$, $\|\bm{x}\|$ denotes its $\ell_2$-norm, and $\mathrm{diag}(\bm{x})$ returns a diagonal matrix with the elements in $\bm{x}$ on its main diagonal. For a matrix $\mathbf{X}$, $\mathrm{Tr}(\mathbf{X})$ and $\mathrm{rank}(\mathbf{X})$ denote its trace and rank, respectively, and $\mathbf{X}\succeq \mathbf{0}$ implies that $\mathbf{X}$ is positive semi-definite. $\mathbf{I}$ and $\mathbf{0}$ denote an identity matrix and an all-zero matrix, respectively, with appropriate dimensions. The distribution of a circularly symmetric complex Gaussian (CSCG) random vector with zero mean and covariance matrix $\bm{\Sigma}$ is denoted by $\mathcal{CN}(\bm{0},\bm{\Sigma})$; and $\sim$ stands for ``distributed as''.

\section{System Model and Problem Formulation}\label{System_model}
As shown in Fig.~\ref{fig_system}, we consider an ISAC system where a dual-functional BS has $N$ RAs deployed in the $x$-$y$ plane to communicate with $K$ users (each equipped with a single isotropic antenna) and to concurrently perform sensing in a given region $\mathcal{A}$ for detecting a potential target. To realize wireless sensing, we assume that the BS can simultaneously transmit and receive the probing signals to estimate the target's position.
For convenience, the sets of RAs and users are denoted by $\mathcal{N}\triangleq \{1,2,\dots,N\}$ and $\mathcal{K}\triangleq \{1,2,\dots,K\}$, respectively. The positions of RA~$n\in \mathcal{N}$ and user~$k\in \mathcal{K}$ are represented by $\mathbf{w}_n \triangleq [w_{n,x},w_{n,y},0]^T$ and $\mathbf{q}_{\mathrm{U},k} \triangleq [q_{\mathrm{U},k,x},q_{\mathrm{U},k,y},q_{\mathrm{U},k,z}]^T$, respectively.
Each RA can independently and flexibly rotate its orientation/boresight to reconfigure the array directional gain pattern to maximize ISAC performance for given spatial distributions of the communication users and the sensing target. Additionally, we assume narrow-band quasi-static channels, i.e., the channel coherence time is sufficiently long to ensure that the wireless environment before and after antenna rotation is approximately unchanged.

To characterize the orientation/boresight of RA~$n$, a pointing vector is defined as~\cite{Wu2024Modeling}
\begin{align}
	\label{deqn_ex1a}
	\vec{\mathbf{f}}_n = [f_{n,x},f_{n,y},f_{n,z}]^T \in \mathbb{R}^{3\times 1},
\end{align}
where $f_{n,x}$, $f_{n,y}$, and $f_{n,z}$ are the projections of $\vec{\mathbf{f}}_n$ on the $x$-, $y$-, and $z$-axes, respectively, and we have $\|\vec{\mathbf{f}}_n\| = 1$ for normalization. 
Moreover, if the reference orientation/boresight of each RA is assumed to be aligned with the positive $z$-axis, the following rotation constraint can be considered to limit the rotational range of each RA and alleviate antenna coupling~\cite{Zheng2025Rotatable}:
\begin{align}
	\label{deqn_ex2a}
	0 \leq \operatorname{arccos}(\vec{\mathbf{f}}_n^T \mathbf{e}_3) \leq \theta_{\mathrm{max}},\; \forall n,
\end{align}
where $\mathbf{e}_3 \triangleq [0,0,1]^T$ and $\theta_{\mathrm{max}} \in [0,\frac{\pi}{2}]$ is the maximum allowable rotation angle for each RA with respect to its reference orientation/boresight.

\subsection{Channel Model}
In order to characterize the effective directional gain, we consider a generic antenna radiation pattern for each RA as follows~\cite{Balanis1996Antenna}:
\begin{equation}
	\label{deqn_ex1b}
	{G_e(\epsilon,\varphi)} =
	\begin{cases}
		{G_{\mathrm{max}} \cos^{2\rho} (\epsilon),}&{\epsilon \in [0,\frac{\pi}{2}), \varphi \in [0,2\pi)} \\
		{0,}&{\text{otherwise},}
	\end{cases}
\end{equation}
where $(\epsilon,\varphi)$ are the elevation and azimuth arrival/departure angles of the signal with respect to the RA's current boresight direction, $\rho \geq 0$ is the directivity factor that characterizes the beamwidth of the mainlobe, and $G_{\mathrm{max}} = 2(2\rho + 1)$ is the maximum gain in the boresight direction (i.e., $\epsilon = 0$) to satisfy the law of power conservation. The directional gain of the radiation pattern in \eqref{deqn_ex1b} depends on the off-boresight angle $\epsilon$ between the signal direction and the antenna boresight direction.

Based on the free space propagation model and the antenna radiation pattern in \eqref{deqn_ex1b}, for a given pointing vector $\vec{\mathbf{f}}_n$, the channel power gain between RA~$n$ and any spatial observation point $\mathbf{q}\in \mathbb{R}^{3\times 1}$ is given by
\begin{align}
	\label{deqn_ex2b}
	g(\vec{\mathbf{f}}_n,\mathbf{q}) = \underbrace{\frac{\beta_0}{d_{n}^2}}_{\mathrm{path}\; \mathrm{gain}} \underbrace{G_{\mathrm{max}} \left[\vec{\mathbf{f}}_n^T \vec{\mathbf{q}}_{n}\right]_{+}^{2\rho}}_{\mathrm{directional}\;\mathrm{gain}},
\end{align}
where $[x]_{+} \triangleq \max \{x,0\}$ denotes the positive operator, $\beta_0 \triangleq \left(\frac{\lambda}{4\pi}\right)^2$ is the channel power gain at a reference distance $d_0 = 1$~meter~(m) with $\lambda$ being the signal wavelength, $d_{n} \triangleq \|\mathbf{q} - \mathbf{w}_n\|$ denotes the distance between RA~$n$ and spatial observation point $\mathbf{q}$, and $\vec{\mathbf{q}}_{n} \triangleq \frac{\mathbf{q} - \mathbf{w}_n}{\|\mathbf{q} - \mathbf{w}_n\|}$ denotes the normalized direction vector from RA~$n$ to spatial observation point $\mathbf{q}$. In addition, $\vec{\mathbf{f}}_n^T \vec{\mathbf{q}}_{n} = \cos(\epsilon_n)$ is the projection between $\vec{\mathbf{f}}_n$ and $\vec{\mathbf{q}}_{n}$. Accordingly, the line-of-sight (LoS) channel between RA~$n$ and this spatial observation point can be expressed as{\footnote{Since ISAC systems usually operate in high-frequency bands, the channels between the BS and users/targets are typically LoS-dominated. Moreover, RA orientation adjustment mainly depends on the dominant propagation direction. Therefore, the LoS channel model is adopted in this paper. For more general propagation environments, multiple LoS/non-LoS components with orientation-dependent gains can be incorporated, while blockage can be modeled through additional attenuation, and sensing clutter can be incorporated as additional scattering components~\cite{Zheng2025Rotatable}.}}
\vspace{-0.2cm}
\begin{align}
	\label{deqn_ex3b}
	h(\vec{\mathbf{f}}_n,\mathbf{q}) = \sqrt{g(\vec{\mathbf{f}}_n,\mathbf{q})} e^{-j\frac{2\pi}{\lambda}d_{n}}.
\end{align}
By stacking the channel coefficients corresponding to all RAs into a vector, the channel between the BS and any spatial observation point can be represented as
\vspace{-0.25cm}
\begin{align}
	\label{deqn_ex4b}
	\mathbf{h}(\mathbf{F},\mathbf{q}) \triangleq \left[h(\vec{\mathbf{f}}_1,\mathbf{q}),h(\vec{\mathbf{f}}_2,\mathbf{q}),\dots,h(\vec{\mathbf{f}}_N,\mathbf{q})\right]^T \in \mathbb{C}^{N\times 1},
\end{align}
where $\mathbf{F} \triangleq [\vec{\mathbf{f}}_1,\vec{\mathbf{f}}_2,\dots,\vec{\mathbf{f}}_N] \in \mathbb{R}^{3\times N}$ denotes the pointing matrix of all RAs.
It is worth noting that antenna rotation affects the effective channel exclusively through the orientation-dependent directional gain in \eqref{deqn_ex2b}. Since the physical positions of the RAs remain unchanged during rotation, neither the propagation distances nor the corresponding phase terms in \eqref{deqn_ex3b} are altered. Therefore, the channel reconfiguration enabled by RA rotation originates from adjusting the directional gains with respect to the relative directions of the communication users and sensing targets.
Based on the channel model defined by \eqref{deqn_ex2b}--\eqref{deqn_ex4b}, the channels for wireless communication and target sensing are presented in the following.

\subsubsection{Communication Channel} The downlink communication channel from the BS to user~$k$ is represented as
\begin{align}
	\label{deqn_ex5b}
	\mathbf{h}_{\mathrm{U}}(\mathbf{F},\mathbf{q}_{\mathrm{U},k}) \triangleq \mathbf{h}(\mathbf{F},\mathbf{q}_{\mathrm{U},k}).
\end{align}
For the considered RA-enabled multi-user communication system, the spatial selectivity introduced by antenna rotation helps to simultaneously enhance the directional gain in multiple intended user directions, thereby mitigating inter-user interference and improving spatial multiplexing gain.

\subsubsection{Round-Trip Sensing Channel} In contrast to communication, since the considered monostatic sensing system relies on the echo signal received at the BS, the round-trip channel is relevant for target sensing. By modeling the target as an unstructured point, the round-trip channel matrix, denoted by $\mathbf{H}_{\mathrm{T}} \in \mathbb{C}^{N\times N}$, for sensing a potential target within sensing region $\mathcal{A}$ can be expressed as
\begin{align}
	\label{deqn_ex6b}
	\mathbf{H}_{\mathrm{T}}(\mathbf{F},\mathbf{q}_{\mathrm{T}}) \triangleq \beta_{\mathrm{T}} \mathbf{h}(\mathbf{F},\mathbf{q}_{\mathrm{T}}) \mathbf{h}^T(\mathbf{F},\mathbf{q}_{\mathrm{T}}),\; \mathbf{q}_{\mathrm{T}}\in \mathcal{A},
\end{align}
where $\mathbf{q}_{\mathrm{T}}$ and $\beta_{\mathrm{T}}$ denote the position and the complex reflection coefficient of the potential target, respectively. We note that for monostatic radar systems, the departure and arrival directions of the probing signal coincide.

\textit{Remark 1:} According to the RA sensing channel model in \eqref{deqn_ex6b}, both its amplitude and phase shift components contain the distance and direction information of the target.
Since the rotation of each RA can be adjusted independently, different antennas may observe the target with different orientations/boresights, thereby enhancing the spatial diversity of the sensing channel. As a result, compared with the conventional fixed-antenna sensing channel, in which all antennas have identical orientations/boresights and thus convey the same directional information, and the isotropic-antenna sensing channel, in which directional information is embedded only in the phase shift component, the considered RA-enabled sensing system is able to capture the distance and direction information of the target from multiple spatial dimensions. 
This richer channel structure is expected to provide additional spatial information for target position estimation. This is quantitatively evaluated based on the CRB derived in Appendix~\ref{CRB} in Section~\ref{Simulation}.

\vspace{-0.2cm}
\subsection{ISAC Model}
In the considered ISAC system, the BS transmits both communication signals and dedicated probing signals to simultaneously perform multi-user downlink communication and target sensing.
To avoid self-interference between communication and sensing at the BS, we assume that they use separate frequency resource blocks, while sharing the same RA orientation configuration.
Despite this resource separation, the communication and sensing functionalities remain coupled through the common RA pointing matrix $\mathbf{F}$, which jointly affects the communication channel in \eqref{deqn_ex5b} and the round-trip sensing channel in \eqref{deqn_ex6b}. Therefore, the communication-sensing trade-off is still governed by the shared RA orientation design.
Let $T$ denote the total number of snapshots of one sensing period. Moreover, we assume that each sensing period does not exceed the length of a coherent time block, during which the channel conditions and the target's geometrical parameters remain approximately unchanged.

\subsubsection{Communication Model} We assume that the communication channel gains $\{\mathbf{h}_{\mathrm{U}}(\mathbf{F},\mathbf{q}_{\mathrm{U},k}), \forall k\in \mathcal{K}\}$ have been acquired by the BS via effective channel estimation methods~\cite{Xiong2025Efficient}. By letting $\mathbf{v}_k\in \mathbb{C}^{N\times 1}$ denote the transmit beamformer for conveying the information symbol to the user~$k$, the received communication signal at user~$k$ in time slot $t$, where $t\in \mathcal{T}\triangleq \{1,2,\dots,T\}$, is expressed as
\vspace{-0.15cm}
\begin{align}
	\label{deqn_ex1c}
	y_k[t] &= \mathbf{h}_{\mathrm{U}}^T(\mathbf{F},\mathbf{q}_{\mathrm{U},k}) \sum_{j=1}^{K}{\mathbf{v}_j c_j[t]} + n_k[t] \nonumber\\
	&=\mathbf{h}_{\mathrm{U}}^T(\mathbf{F},\mathbf{q}_{\mathrm{U},k}) \mathbf{v}_k c_k[t] \hspace{-0.05cm}+\hspace{-0.05cm} \sum_{j\ne k}{\mathbf{h}_{\mathrm{U}}^T(\mathbf{F},\mathbf{q}_{\mathrm{U},k}) \mathbf{v}_j c_j[t]} \hspace{-0.05cm}+\hspace{-0.05cm} n_k[t],
\end{align}
where $c_k[t]$ denotes the independent and identically distributed (i.i.d.) information-bearing symbol for user~$k$ with $\mathbb{E}\left[|c_k[t]|^2\right] = 1$, and $n_k[t] \sim \mathcal{CN}(0,\sigma_k^2)$ is the complex additive white Gaussian noise (AWGN) at user $k$ with zero mean and variance $\sigma_k^2$. Accordingly, the receive signal-to-interference-plus-noise ratio (SINR) for user~$k$ is given by
\vspace{-0.1cm}
\begin{align}
	\label{deqn_ex2c}
	\gamma_k(\mathbf{F},\mathbf{V}) = \frac{|\mathbf{h}_{\mathrm{U}}^T(\mathbf{F},\mathbf{q}_{\mathrm{U},k}) \mathbf{v}_k|^2}{\sum_{j\ne k}{|\mathbf{h}_{\mathrm{U}}^T(\mathbf{F},\mathbf{q}_{\mathrm{U},k}) \mathbf{v}_j|^2} + \sigma_k^2},
\end{align}
where $\mathbf{V}\triangleq \left[\mathbf{v}_1,\mathbf{v}_2,\dots,\mathbf{v}_K \right]\in \mathbb{C}^{N\times K}$ denotes the transmit beamforming matrix. Then, the achievable rate of user~$k$ is given by
\vspace{-0.15cm}
\begin{align}
	\label{deqn_ex3c}
	R_k(\mathbf{F},\mathbf{V}) = \operatorname{log}_2 \left(1 + \gamma_k(\mathbf{F},\mathbf{V})\right).
\end{align}

\subsubsection{Sensing Model} For any given RA pointing matrix $\mathbf{F}$, the received echo signal from potential target $\mathbf{q}_{\mathrm{T}}\in \mathcal{A}$ in time slot $t$ can be expressed as
\begin{align}
	\label{deqn_ex4c}
	\mathbf{y}_{\mathrm{s}}[t] = \mathbf{H}_{\mathrm{T}}(\mathbf{F},\mathbf{q}_{\mathrm{T}}) \mathbf{s}[t] + \mathbf{n}_{\mathrm{s}}[t],
\end{align}
where $\mathbf{s}[t]\in \mathbb{C}^{N\times 1}$ denotes the transmit probing signal from the BS, and $\mathbf{n}_{\mathrm{s}}[t] \sim \mathcal{CN}(\mathbf{0},\sigma_{\mathrm{s}}^2\mathbf{I}_N)$ is the AWGN vector at the BS with variance $\sigma_{\mathrm{s}}^2$. Accordingly, the received echo signal power from the target at the BS is given by
\begin{align}
	\label{deqn_ex5c}
	\hspace{-0.15cm}P(\mathbf{F},\mathbf{S},\mathbf{q}_{\mathrm{T}}) =& \mathbb{E}\left\{\|\beta_{\mathrm{T}} \mathbf{h}(\mathbf{F},\mathbf{q}_{\mathrm{T}}) \mathbf{h}^T(\mathbf{F},\mathbf{q}_{\mathrm{T}}) \mathbf{s}[t]\|^2\right\} \nonumber\\
	=& |\beta_{\mathrm{T}}|^2 \|\mathbf{h}(\mathbf{F},\mathbf{q}_{\mathrm{T}})\|^2 \mathbf{h}^T(\mathbf{F},\mathbf{q}_{\mathrm{T}}) \mathbf{S} \mathbf{h}^{\ast}(\mathbf{F},\mathbf{q}_{\mathrm{T}}),\hspace{-0.1cm}
\end{align}
where $\mathbf{S}\triangleq \mathbb{E}\left\{\mathbf{s}[t]\mathbf{s}^H[t] \right\}\in \mathbb{C}^{N\times N}$ with $\mathbf{S} \succeq \mathbf{0}$ denotes the covariance matrix of the dedicated probing signal. 
Since the echo signal power is a practically measurable metric that directly reflects the received sensing signal strength, we adopt it for evaluating the sensing performance. Since the noise variance $\sigma_{\mathrm{s}}^2$ is fixed in the considered sensing model, the sensing signal-to-noise ratio (SNR) is proportional to the received echo signal power $P(\mathbf{F},\mathbf{S},\mathbf{q}_{\mathrm{T}})$. Therefore, maximizing the echo signal power directly improves the received sensing SNR. For detection, a higher sensing SNR generally improves the detection probability under a fixed false-alarm probability. For estimation and localization, a higher sensing SNR generally provides more informative observations and tends to improve estimation performance, characterized, for example, by the CRB. Therefore, although the echo signal power does not directly characterize detection probability or localization error, it serves as a physically meaningful metric for the considered regional sensing task.

\subsection{Problem Formulation}
To balance the trade-off between communication and sensing performance and to achieve uniform sensing performance across the entire target region, in this paper, we aim to jointly optimize the RA pointing matrix $\mathbf{F}$, the transmit beamforming matrix $\mathbf{V}$, and the probing-signal covariance matrix $\mathbf{S}$ for maximization of the minimum received echo signal power of the target within the sensing region $\mathcal{A}$, while guaranteeing that the communication rate of each user does not fall below a given threshold. The resulting optimization problem can be formulated as follows:
\begin{subequations}\label{eq:0}
	\begin{alignat}{2}
		(\mathrm{P1}):\; \max_{\mathbf{F},\mathbf{V},\mathbf{S}\succeq \mathbf{0}} \quad & \min_{\mathbf{q}_{\mathrm{T}}\in \mathcal{A}}{P(\mathbf{F},\mathbf{S},\mathbf{q}_{\mathrm{T}})} & \label{eq:0A}\\
		\mbox{s.t.} \quad
		& R_k(\mathbf{F},\mathbf{V}) \geq R_{\mathrm{min},k},\; \forall k, \label{eq:0B}\\
		& \sum_{k=1}^{K}{\|\mathbf{v}_k\|^2} \leq P_{\mathrm{max,c}}, \label{eq:0C}\\
		& \operatorname{Tr}(\mathbf{S}) \leq P_{\mathrm{max,s}}, \label{eq:0D}\\
		& 0 \leq \arccos(\vec{\mathbf{f}}_n^T \mathbf{e}_3) \leq \theta_{\mathrm{max}},\; \forall n, \label{eq:0E}\\
		& \|\vec{\mathbf{f}}_n\|^2 = 1, \; \forall n, \label{eq:0F}
	\end{alignat}
\end{subequations}
where constraint \eqref{eq:0B} ensures that user~$k$ achieves its required minimum communication rate $R_{\mathrm{min},k}$; constraints \eqref{eq:0C} and \eqref{eq:0D} limit the average transmit powers for communication and sensing to $P_{\mathrm{max,c}}$ and $P_{\mathrm{max,s}}$, respectively; constraint \eqref{eq:0E} constrains the rotation of each RA to remain within the allowable rotational range; constraint \eqref{eq:0F} ensures that $\vec{\mathbf{f}}_n$ is a unit-norm vector.
Note that, given the geometric information of the sensing region of interest, the round-trip sensing channel associated with any spatial point in the region can be determined according to \eqref{deqn_ex6b}. Therefore, the echo signal power over the entire region can be represented as a function of $\mathbf{F}$ and $\mathbf{S}$ as shown in \eqref{eq:0A}, without requiring point-by-point channel estimation.
In the following sections, we provide the solutions to problem (P1) for both the single-user and point-target case and the multi-user and regional sensing case.

\section{Single-User and Point-Target Case}\label{Single_user}
In this section, we consider the special case of single-user communication and point-target sensing, i.e., $K = 1$ and the sensing region $\mathcal{A}$ reduces to a single point, to gain essential insights from problem (P1). In this case, by denoting the locations of the user and the target by $\mathbf{q}_{\mathrm{U}}$ and $\mathbf{q}_{\mathrm{T}}$, respectively, and dropping the user index for simplicity, problem (P1) simplifies to
\begin{subequations}\label{eq:s1}
	\begin{alignat}{2}
		(\mathrm{P2}):\; \max_{\mathbf{F},\mathbf{v},\mathbf{S}\succeq \mathbf{0}} \quad & \|\mathbf{h}(\mathbf{F},\mathbf{q}_{\mathrm{T}})\|^2 \mathbf{h}^T(\mathbf{F},\mathbf{q}_{\mathrm{T}}) \mathbf{S} \mathbf{h}^{\ast}(\mathbf{F},\mathbf{q}_{\mathrm{T}}) & \label{eq:s1A}\\
		\mbox{s.t.} \quad
		& \frac{1}{\sigma^2} \left|\mathbf{h}^T(\mathbf{F},\mathbf{q}_{\mathrm{U}})\mathbf{v}\right|^2 \geq \bar{\gamma}, \label{eq:s1B}\\
		& \|\mathbf{v}\|^2 \leq P_{\mathrm{max,c}}, \label{eq:s1C}\\
		& \eqref{eq:0D}-\eqref{eq:0F}, \label{eq:s1D}
	\end{alignat}
\end{subequations}
where $\bar{\gamma} \triangleq 2^{R_{\mathrm{min}}} - 1$ denotes the required minimum SNR of the user.

Recall that communication and sensing employ separate frequency resource blocks and are subject to separate transmit-power constraints. Thus, for any given RA pointing matrix $\mathbf{F}$, the communication beamforming vector $\mathbf{v}$ and the probing-signal covariance matrix $\mathbf{S}$ can be optimized independently. For the considered single-user case, there is no inter-user interference, and maximum-ratio transmission (MRT) is therefore optimal for maximizing the communication SNR on the left hand side of constraint \eqref{eq:s1B}.
Accordingly, the optimal $\mathbf{v}$ for problem (P2) is given by $\mathbf{v}^{\star} = \sqrt{P_{\mathrm{max,c}}} \frac{\mathbf{h}^{\ast}(\mathbf{F},\mathbf{q}_{\mathrm{U}})}{\|\mathbf{h}(\mathbf{F},\mathbf{q}_{\mathrm{U}})\|}$.
Furthermore, for any given $\mathbf{F}$, maximizing the objective function in \eqref{eq:s1A} with respect to $\mathbf{S}$ is equivalent to maximizing $\mathbf{h}^T(\mathbf{F},\mathbf{q}_{\mathrm{T}}) \mathbf{S} \mathbf{h}^{\ast}(\mathbf{F},\mathbf{q}_{\mathrm{T}})$ subject to $\mathbf{S}\succeq \mathbf{0}$ and \eqref{eq:0D}. Thus, the optimal probing-signal covariance matrix allocates the entire sensing power to the dominant spatial mode $\mathbf{h}^{\ast}(\mathbf{F},\mathbf{q}_{\mathrm{T}})$, yielding $\mathbf{S}^{\star} = P_{\mathrm{max,s}} \frac{\mathbf{h}^{\ast}(\mathbf{F},\mathbf{q}_{\mathrm{T}}) \mathbf{h}^T(\mathbf{F},\mathbf{q}_{\mathrm{T}})}{\|\mathbf{h}(\mathbf{F},\mathbf{q}_{\mathrm{T}})\|^2}$.
Substituting $\mathbf{v}^{\star}$ and $\mathbf{S}^{\star}$ into problem (P2) yields the following problem:
\begin{subequations}\label{eq:s2}
	\begin{alignat}{2}
		(\mathrm{P3}):\; \max_{\mathbf{F}} \quad & \left(\sum_{n=1}^{N}{\frac{1}{d_{{\mathrm{T}},n}^2}\left[\vec{\mathbf{f}}_n^T \vec{\mathbf{q}}_{{\mathrm{T}},n}\right]_{+}^{2\rho}} \right)^2 & \label{eq:s2A}\\
		\mbox{s.t.} \quad
		& \sum_{n=1}^{N}{\frac{1}{d_{{\mathrm{U}},n}^2}\left[\vec{\mathbf{f}}_n^T \vec{\mathbf{q}}_{{\mathrm{U}},n}\right]_{+}^{2\rho}} \geq \omega, \label{eq:s2B}\\
		& \eqref{eq:0E}, \eqref{eq:0F}, \label{eq:s2C}
	\end{alignat}
\end{subequations}
where $\omega \triangleq \frac{\bar{\gamma}\sigma^2}{\beta_0 G_{\mathrm{max}} P_{\mathrm{max,c}}}$ and the constant term is omitted in the objective function \eqref{eq:s2A}. In addition, $d_{\mathrm{U},n} = \|\mathbf{q}_{\mathrm{U}} - \mathbf{w}_n\|$ and $d_{\mathrm{T},n} = \|\mathbf{q}_{\mathrm{T}} - \mathbf{w}_n\|$ denote the distance between RA~$n$ and the user and that between RA~$n$ and the target, respectively, and $\vec{\mathbf{q}}_{{\mathrm{U}},n} = \frac{\mathbf{q}_{{\mathrm{U}}} - \mathbf{w}_n}{\|\mathbf{q}_{{\mathrm{U}}} - \mathbf{w}_n\|}$ and $\vec{\mathbf{q}}_{{\mathrm{T}},n} = \frac{\mathbf{q}_{{\mathrm{T}}} - \mathbf{w}_n}{\|\mathbf{q}_{{\mathrm{T}}} - \mathbf{w}_n\|}$ denote the normalized direction vectors from RA~$n$ to the user and the target, respectively. Due to the non-convexity of \eqref{eq:0F}, \eqref{eq:s2A}, and \eqref{eq:s2B}, problem (P3) is difficult to solve directly. 

To obtain a tractable RA pointing design and to facilitate the subsequent analysis, we first simplify the communication constraint \eqref{eq:s2B}.
Since $x^{\frac{1}{2\rho}}$ is concave for $\rho \geq \frac{1}{2}$, based on Jensen's inequality~\cite{Boyd2004Convex}, a lower bound for the left-hand side of \eqref{eq:s2B} can be obtained as
\begin{align}
	\label{deqn_ex1d}
	\hspace{-0.4cm}\left(\sum_{n=1}^{N}{\frac{[\vec{\mathbf{f}}_n^T \vec{\mathbf{q}}_{{\mathrm{U}},n}]_{+}^{2\rho}}{d_{{\mathrm{U}},n}^2}} \right)^{\frac{1}{2\rho}} \hspace{-0.15cm}\geq\hspace{-0.05cm} N^{\frac{1-2\rho}{2\rho}}\sum_{n=1}^{N}{\left(\frac{[\vec{\mathbf{f}}_n^T \vec{\mathbf{q}}_{{\mathrm{U}},n}]_{+}^{2\rho}}{d_{{\mathrm{U}},n}^2}\right)^{\frac{1}{2\rho}}}\hspace{-0.1cm}, \hspace{-0.4cm}
\end{align}
where equality holds if and only if the following condition is satisfied:
\begin{align}
	\label{deqn_ex2d}
	\frac{[\vec{\mathbf{f}}_1^T \vec{\mathbf{q}}_{{\mathrm{U}},1}]_{+}^{2\rho}}{d_{{\mathrm{U}},1}^2} = \frac{[\vec{\mathbf{f}}_2^T \vec{\mathbf{q}}_{{\mathrm{U}},2}]_{+}^{2\rho}}{d_{{\mathrm{U}},2}^2} =\dots= \frac{[\vec{\mathbf{f}}_N^T \vec{\mathbf{q}}_{{\mathrm{U}},N}]_{+}^{2\rho}}{d_{{\mathrm{U}},N}^2}.
\end{align}
Therefore, using \eqref{deqn_ex1d}, communication constraint \eqref{eq:s2B} can be approximated as
\begin{align}
	\label{deqn_ex3d}
	\sum_{n=1}^{N}{\frac{[\vec{\mathbf{f}}_n^T \vec{\mathbf{q}}_{{\mathrm{U}},n}]_{+}}{d_{{\mathrm{U}},n}^{1/\rho}} } \geq N^{\frac{2\rho-1}{2\rho}}\omega^{\frac{1}{2\rho}}.
\end{align}
In addition, since the summation term in \eqref{eq:s2A} is nonnegative, removing its outer square does not change the optimal pointing matrix $\mathbf{F}$. Thus, replacing \eqref{eq:s2B} with \eqref{deqn_ex3d} yields the following conservative approximation of (P3):
\begin{subequations}\label{eq:s6}
	\begin{alignat}{2}
		(\mathrm{P4}):\; \max_{\mathbf{F}} \quad & \sum_{n=1}^{N}{\frac{1}{d_{{\mathrm{T}},n}^2}\left[\vec{\mathbf{f}}_n^T \vec{\mathbf{q}}_{{\mathrm{T}},n}\right]_{+}^{2\rho}} & \label{eq:s6A}\\
		\mbox{s.t.} \quad
		& \eqref{eq:0E}, \eqref{eq:0F}, \eqref{deqn_ex3d}. \label{eq:s6B}
	\end{alignat}
\end{subequations}

Furthermore, since the left-hand side of \eqref{deqn_ex3d} consists of $N$ nonnegative per-RA terms, constraint \eqref{deqn_ex3d} is guaranteed if each term satisfies
\begin{align}
	\label{deqn_ex4d}
	\frac{[\vec{\mathbf{f}}_n^T \vec{\mathbf{q}}_{{\mathrm{U}},n}]_{+}}{d_{{\mathrm{U}},n}^{1/\rho}}\geq \left(\frac{\omega}{N}\right)^{\frac{1}{2\rho}},\; \forall n,
\end{align}
which constitutes a sufficient condition for \eqref{deqn_ex3d}. Indeed, according to \eqref{deqn_ex1d} and \eqref{deqn_ex2d}, summing the above inequalities over all RAs yields exactly the lower bound required in \eqref{deqn_ex3d}.
With the aggregate communication constraint \eqref{deqn_ex3d} replaced by these per-RA constraints in \eqref{deqn_ex4d}, the pointing vectors are decoupled across the RAs. Since the objective function in \eqref{eq:s6A} is also separable with respect to the pointing vectors, the resulting optimization problem can be decomposed into $N$ independent per-RA subproblems. For RA~$n$, the corresponding subproblem is given by
\begin{subequations}\label{eq:s7}
	\begin{alignat}{2}
		(\mathrm{P5}):\; \max_{\vec{\mathbf{f}}_n} \quad & \vec{\mathbf{f}}_n^T \vec{\mathbf{q}}_{{\mathrm{T}},n} & \label{eq:s7A}\\
		\mbox{s.t.} \quad
		& \vec{\mathbf{f}}_n^T \vec{\mathbf{q}}_{{\mathrm{U}},n}\geq \cos \theta_{\mathrm{U},n}, \label{eq:s7B}\\
		& \vec{\mathbf{f}}_n^T \mathbf{e}_3 \geq \cos \theta_{\mathrm{max}}, \label{eq:s7C}\\
		& \|\vec{\mathbf{f}}_n\|^2 = 1, \label{eq:s7D}
	\end{alignat}
\end{subequations}
where constraint \eqref{eq:s7B} is obtained from \eqref{deqn_ex4d}, with $\theta_{\mathrm{U},n}\triangleq \arccos \left(\left(\frac{\omega}{N}d_{{\mathrm{U}},n}^2 \right)^{\frac{1}{2\rho}}\right)$ denoting the maximum allowable angle between the pointing vector $\vec{\mathbf{f}}_n$ and the user direction $\vec{\mathbf{q}}_{{\mathrm{U}},n}$ under the corresponding per-RA communication requirement. Clearly, $\theta_{\mathrm{U},n}$ increases as $R_{\mathrm{min}}$ decreases. Constraint \eqref{eq:s7C} is equivalent to the rotation constraint \eqref{eq:0E}.
It should be emphasized that (P5) is derived based on the sufficient conditions above. Hence, its closed-form solution is optimal for (P5), but is not necessarily globally optimal for the original problem (P3).
In addition, according to the optimal RA pointing design for the single-user free-space communication system derived in \cite{Wu2024Modeling}, the maximum achievable user rate is
\begin{align}
	\label{deqn_ex5d}
	R_{\mathrm{min}}\leq \operatorname{log}_2\left(1 + \frac{P_{\mathrm{max,c}}}{\sigma^2}\sum_{n = 1}^{N}{G_{\mathrm{max}} \left(\frac{\lambda}{4\pi d_{\mathrm{U},n}}\right)^2}\right).
\end{align}
Thus, the maximum feasible value of $\omega$ is given by $\sum_{n = 1}^{N}{d_{\mathrm{U},n}^{-2}}$. When $d_{\mathrm{U},1}\approx d_{\mathrm{U},2}\approx \dots \approx d_{\mathrm{U},N}$, this ensures $0< \left(\frac{\omega}{N}d_{\mathrm{U},n}^2 \right)^{\frac{1}{2\rho}} \leq 1$ and hence $\theta_{\mathrm{U},n}\in [0,\frac{\pi}{2}]$.

\textit{Remark 2:} If both the user and the target are located in the far-field region of the BS, the signal transmission between the BS and the user/target can be characterized by a plane-wave model. Thus, the distance and direction of the user/target relative to all RAs on the array are approximately the same. In this case, we have $d_{{\mathrm{X}},1}\approx d_{{\mathrm{X}},2}\approx \dots \approx d_{{\mathrm{X}},N}$ and $\vec{\mathbf{q}}_{{\mathrm{X}},1}\approx \vec{\mathbf{q}}_{{\mathrm{X}},2}\approx \dots \approx \vec{\mathbf{q}}_{{\mathrm{X}},N}$ with $\mathrm{X}\in \{\mathrm{U},\mathrm{T}\}$.
Accordingly, $\theta_{\mathrm{U},n}$ in \eqref{eq:s7B} is also approximately identical across the RAs, such that the per-RA subproblems in (P5) can be approximated by the same pointing optimization problem. Therefore, a common pointing vector can be adopted for all RAs under the far-field approximation, i.e., one may choose $\vec{\mathbf{f}}_1 = \vec{\mathbf{f}}_2 =\dots = \vec{\mathbf{f}}_N$.

\begin{figure}[!t]
	\centering
	\subfloat[User-target sector spanned by $\vec{\mathbf{q}}_{\mathrm{U},n}$ and $\vec{\mathbf{q}}_{\mathrm{T},n}$]{
		\includegraphics[width=1.5in]{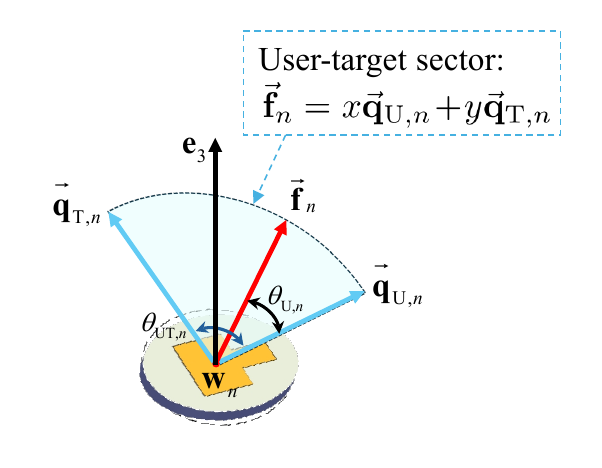}}\hspace{+0.4cm}
	\subfloat[Boundary of the allowable rotation cone]{
		\includegraphics[width=1.5in]{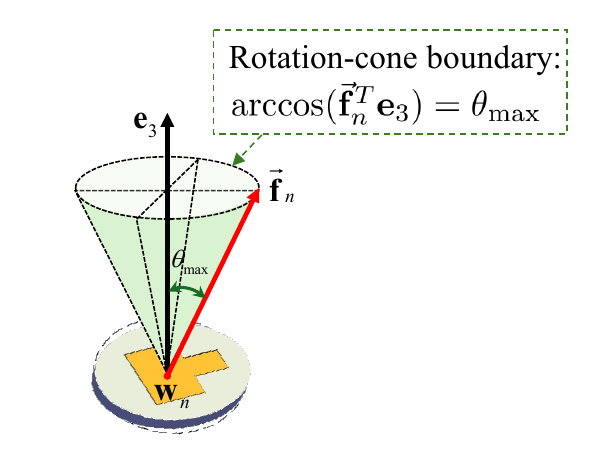}}
	\caption{Illustration of the two geometric cases for the optimal pointing vector of RA~$n$.}
	\label{fig_sector_cone}\vspace{-0.3cm}
\end{figure}

For problem (P5), constraint \eqref{eq:s7B} specifies that pointing vector $\vec{\mathbf{f}}_n$ must lie within a spherical cone on the unit sphere, with the user direction $\vec{\mathbf{q}}_{\mathrm{U},n}$ as its axis and a half-angle of $\theta_{\mathrm{U},n}$. Meanwhile, constraint \eqref{eq:s7C} requires that $\vec{\mathbf{f}}_n$ lies within another spherical cone on the unit sphere with the $z$-axis (i.e., $\mathbf{e}_3$) as its axis and a half-angle of $\theta_{\mathrm{max}}$.
The two spherical cones can intersect only if the angle between their axes is not greater than the sum of their half-angles. This condition can be expressed as $\theta_{\mathrm{max}} + \theta_{\mathrm{U},n} \geq \psi_{\mathrm{U},n}$, where $\psi_{\mathrm{U},n}\triangleq \arccos(\vec{\mathbf{q}}_{\mathrm{U},n}^T \mathbf{e}_3)$ denotes the angle between the user direction $\vec{\mathbf{q}}_{\mathrm{U},n}$ and the $z$-axis. This condition must be satisfied to ensure the existence of feasible solutions to problem (P5).
Geometrically, problem (P5) seeks a unit-norm pointing vector that maximizes its alignment with the target direction $\vec{\mathbf{q}}_{\mathrm{T},n}$ over the intersection of the communication-feasible cone defined by \eqref{eq:s7B} and the rotation-feasible cone defined by \eqref{eq:s7C}.
Therefore, the optimal pointing vector $\vec{\mathbf{f}}_n$ can either lie within the sector spanned by $\vec{\mathbf{q}}_{\mathrm{U},n}$ and $\vec{\mathbf{q}}_{\mathrm{T},n}$ (i.e., $\vec{\mathbf{f}}_n = x\vec{\mathbf{q}}_{\mathrm{U},n} + y\vec{\mathbf{q}}_{\mathrm{T},n}$ with $x,y\geq 0$ and $\|x\vec{\mathbf{q}}_{\mathrm{U},n} + y\vec{\mathbf{q}}_{\mathrm{T},n}\| = 1$), as shown in Fig.~\ref{fig_sector_cone}(a), or on the boundary of the allowable rotation cone defined by \eqref{eq:s7C} (i.e., $\operatorname{arccos}(\vec{\mathbf{f}}_n^T \mathbf{e}_3) = \theta_{\mathrm{max}}$), as shown in Fig.~\ref{fig_sector_cone}(b).
For any other choice, we can always find a unit-norm vector that achieves a larger value of $\vec{\mathbf{f}}_n^T \vec{\mathbf{q}}_{\mathrm{T},n}$. In the following, we present the solutions to problem (P5) corresponding to these two cases.

\subsection{Solution within the User-Target Sector}\label{Planer_sector}
By relaxing constraint \eqref{eq:s7C}, or equivalently setting $\theta_{\mathrm{max}} = \frac{\pi}{2}$, we first consider the solution within the sector spanned by $\vec{\mathbf{q}}_{\mathrm{U},n}$ and $\vec{\mathbf{q}}_{\mathrm{T},n}$. Problem (P5) then reduces to finding a unit-norm pointing vector that is as closely aligned with the target direction $\vec{\mathbf{q}}_{\mathrm{T},n}$ as possible while satisfying the communication constraint \eqref{eq:s7B}. 
Since both $\vec{\mathbf{f}}_n$ and $\vec{\mathbf{q}}_{\mathrm{T},n}$ are unit-norm vectors, maximizing $\vec{\mathbf{f}}_n^T\vec{\mathbf{q}}_{\mathrm{T},n}$ is equivalent to minimizing their angular separation.
Let $\theta_{\mathrm{UT},n}\triangleq \arccos \left(\vec{\mathbf{q}}_{\mathrm{U},n}^T \vec{\mathbf{q}}_{\mathrm{T},n}\right)$ denote the angle between the user and target directions.
Constraint \eqref{eq:s7B} restricts the angular deviation of $\vec{\mathbf{f}}_n$ from $\vec{\mathbf{q}}_{\mathrm{U},n}$ to no more than $\theta_{\mathrm{U},n}$.
Therefore, if $\theta_{\mathrm{U},n}\geq \theta_{\mathrm{UT},n}$, the target direction itself is feasible and the optimal pointing vector is $\vec{\mathbf{f}}_n^{\star} = \vec{\mathbf{q}}_{\mathrm{T},n}$.
Otherwise, the endpoint of the optimal pointing vector lies on the shortest geodesic arc on the unit sphere from $\vec{\mathbf{q}}_{\mathrm{U},n}$ toward $\vec{\mathbf{q}}_{\mathrm{T},n}$, at the boundary of the communication constraint.
Accordingly, the optimal angular displacement from the user direction is $\alpha_{n}^{\star} = \min \{\theta_{\mathrm{U},n},\theta_{\mathrm{UT},n}\}$. Define the corresponding normalized angular displacement as
\begin{align}
	\label{deqn_ex0g}
	\varrho_n \triangleq \frac{\alpha_{n}^{\star}}{\theta_{\mathrm{UT},n}} = \min \left\{\frac{\theta_{\mathrm{U},n}}{\theta_{\mathrm{UT},n}}, 1\right\}.
\end{align}
For $0 < \theta_{\mathrm{UT},n} < \pi$, spherical linear interpolation (Slerp) along the above geodesic arc gives the optimal solution to problem (P5) as~\cite{Dam1998Quaternions}
\begin{align}
	\label{deqn_ex1g}
	\vec{\mathbf{f}}_n^{\star} = \frac{\sin \left((1-\varrho_n)\theta_{\mathrm{UT},n}\right)}{\sin \theta_{\mathrm{UT},n}}\vec{\mathbf{q}}_{\mathrm{U},n} + \frac{\sin \left(\varrho_n \theta_{\mathrm{UT},n}\right)}{\operatorname{sin}\theta_{\mathrm{UT},n}}\vec{\mathbf{q}}_{\mathrm{T},n},
\end{align}
For the special case $\theta_{\mathrm{UT},n} = 0$, the user and target directions coincide and we directly have $\vec{\mathbf{f}}_n^{\star} = \vec{\mathbf{q}}_{\mathrm{U},n} = \vec{\mathbf{q}}_{\mathrm{T},n}$.
Substituting \eqref{deqn_ex1g} into the objective function \eqref{eq:s7A}, the maximum achievable projection toward the target direction is
\begin{equation}
	\label{deqn_ex2g}
	\max_{\vec{\mathbf{f}}_n\in\mathcal{F}_n}{\vec{\mathbf{f}}_n^T\vec{\mathbf{q}}_{\mathrm{T},n}} =
	\begin{cases}
		{1,}&{\theta_{\mathrm{U},n}\geq \theta_{\mathrm{UT},n}} \\
		{\cos(\theta_{\mathrm{UT},n} -\theta_{\mathrm{U},n}),}&{\theta_{\mathrm{U},n} < \theta_{\mathrm{UT},n},}
	\end{cases}\hspace{-0.25cm}
\end{equation}
where $\mathcal{F}_n$ denotes the feasible set of problem (P5).

Equation \eqref{deqn_ex2g} provides a clear geometric interpretation of the communication-sensing trade-off. When $\theta_{\mathrm{U},n}\geq \theta_{\mathrm{UT},n}$, the RA can directly point toward the target without violating the communication requirement, yielding the maximum projection of one. In contrast, when $\theta_{\mathrm{U},n}< \theta_{\mathrm{UT},n}$, the communication constraint becomes active and prevents the RA from fully aligning with the target direction. In this case, the optimal pointing vector lies on the boundary of the communication-feasible region and is rotated from the user direction toward the target by $\theta_{\mathrm{U},n}$.
Since $\theta_{\mathrm{U},n}$ increases as the required minimum communication rate $R_{\min}$ decreases, relaxing the communication requirement enlarges the allowable angular deviation from the user direction. Consequently, the RA can point closer to $\vec{\mathbf{q}}_{\mathrm{T},n}$, increasing $\vec{\mathbf{f}}_n^T\vec{\mathbf{q}}_{\mathrm{T},n}$ and hence the received echo signal power. This reveals that a less stringent communication requirement provides greater spatial flexibility for sensing in RA-enabled ISAC systems.

\subsection{Solution on the Boundary of the Rotation Cone}
If the candidate pointing vector $\vec{\mathbf{f}}_n$ obtained from \eqref{deqn_ex1g} violates rotation constraint \eqref{eq:s7C}, the rotation constraint becomes active at the optimum. Consequently, the optimal pointing vector lies on the boundary of the allowable rotation cone specified by \eqref{eq:s7C}, as illustrated in Fig.~\ref{fig_sector_cone}(b). In other words, the angle between the optimal pointing vector and the $z$-axis is $\theta_{\mathrm{max}}$. Accordingly, problem (P5) can be transformed into the following problem:
\begin{subequations}\label{eq:s4}
	\begin{alignat}{2}
		(\mathrm{P6}):\; \max_{\phi_n,\vec{\mathbf{f}}_n} \quad & \vec{\mathbf{f}}_n^T \vec{\mathbf{q}}_{\mathrm{T},n} & \label{eq:s4A}\\
		\mbox{s.t.} \quad
		& \operatorname{cos}\theta_{\mathrm{U},n}\leq \vec{\mathbf{f}}_n^T \vec{\mathbf{q}}_{\mathrm{U},n} \leq 1, \label{eq:s4B}\\
		& \vec{\mathbf{f}}_n = \begin{bmatrix}
			\sin\theta_{\mathrm{max}}\cos\phi_n\\
			\sin\theta_{\mathrm{max}}\sin\phi_n\\
			\cos\theta_{\mathrm{max}}
		\end{bmatrix},\label{eq:s4C}\\
		& 0 \leq \phi_n \leq 2\pi, \label{eq:s4D}
	\end{alignat}
\end{subequations}
where constraint \eqref{eq:s4B} is equivalent to \eqref{eq:s7B}, and constraints \eqref{eq:s4C} and \eqref{eq:s4D} jointly define a unit-norm vector whose angle with the $z$-axis is $\theta_{\mathrm{max}}$. Specifically, $\phi_n$ denotes the azimuth angle of $\vec{\mathbf{f}}_n$ in the $x$-$y$ plane.
Let $\phi_{\mathrm{U},n}\triangleq \operatorname{arctan2}\left(\vec{\mathbf{q}}_{\mathrm{U},n}^T\mathbf{e}_2, \vec{\mathbf{q}}_{\mathrm{U},n}^T\mathbf{e}_1\right)$ denote the azimuth angle of the user direction $\vec{\mathbf{q}}_{\mathrm{U},n}$, where $\mathbf{e}_1 = [1,0,0]^T$, $\mathbf{e}_2 = [0,1,0]^T$, and $\operatorname{arctan2}(x,y)$ is the four-quadrant inverse tangent function determined by the sine component $x$ and the cosine component $y$. Then, $\vec{\mathbf{q}}_{\mathrm{U},n}$ can be expressed as $\vec{\mathbf{q}}_{\mathrm{U},n} = [\sin\psi_{\mathrm{U},n} \cos\phi_{\mathrm{U},n},\sin\psi_{\mathrm{U},n} \sin\phi_{\mathrm{U},n},\cos\psi_{\mathrm{U},n}]^T$. Accordingly, constraint \eqref{eq:s4B} can be rewritten as
\begin{align}
	\label{deqn_ex3g}
	\cos(\phi_n - \phi_{\mathrm{U},n}) &\geq \left[\frac{\cos\theta_{\mathrm{U},n} - \cos\theta_{\mathrm{max}}\cos\psi_{\mathrm{U},n}}{\sin\theta_{\mathrm{max}}\sin\psi_{\mathrm{U},n}}\right]_{-1}^{1}\nonumber\\
	&\triangleq \operatorname{cos}\left(\Delta\phi_n\right),
\end{align}
where $[x]_a^b\triangleq \operatorname{min}\{\operatorname{max}\{x,a\},b\}$.
Under the feasibility condition $\theta_{\mathrm{U},n} \geq \psi_{\mathrm{U},n} - \theta_{\mathrm{max}}$, we have $\cos\theta_{\mathrm{U},n} - \cos\theta_{\mathrm{max}}\cos\psi_{\mathrm{U},n} - \sin\theta_{\mathrm{max}}\sin\psi_{\mathrm{U},n} = \cos\theta_{\mathrm{U},n} - \cos(\theta_{\mathrm{max}} - \psi_{\mathrm{U},n}) \leq 0$. It then follows that $\frac{\cos\theta_{\mathrm{U},n} - \cos\theta_{\mathrm{max}}\cos\psi_{\mathrm{U},n}}{\sin\theta_{\mathrm{max}}\sin\psi_{\mathrm{U},n}} \leq 1$, and thus constraint~\eqref{deqn_ex3g} is feasible. This implies that the azimuth-angle difference between user direction $\vec{\mathbf{q}}_{\mathrm{U},n}$ and pointing vector $\vec{\mathbf{f}}_n$ cannot exceed $\Delta\phi_n$ if constraint~\eqref{eq:s4B} is to be satisfied. Considering the geometric relationship between vectors $\vec{\mathbf{f}}_n$, $\vec{\mathbf{q}}_{\mathrm{U},n}$, and $\vec{\mathbf{q}}_{\mathrm{T},n}$, problem (P6) can be equivalently transformed into
\begin{subequations}\label{eq:s5}
	\begin{alignat}{2}
		(\mathrm{P7}): \min_{\phi_n} \quad & |\phi_{\mathrm{T},n} - \phi_n| & \label{eq:s5A}\\
		\mbox{s.t.} \quad
		& |\phi_{\mathrm{U},n} - \phi_n|\leq \Delta\phi_n, \label{eq:s5B}\\
		& 0 \leq \phi_n \leq 2\pi, \label{eq:s5C}
	\end{alignat}
\end{subequations}
where $\phi_{\mathrm{T},n}\triangleq \operatorname{arctan2}\left(\vec{\mathbf{q}}_{\mathrm{T},n}^T\mathbf{e}_2, \vec{\mathbf{q}}_{\mathrm{T},n}^T\mathbf{e}_1\right)$ denotes the azimuth angle of the target direction $\vec{\mathbf{q}}_{\mathrm{T},n}$ in the $x$-$y$ plane, and constraint \eqref{eq:s5B} is obtained from \eqref{deqn_ex3g} and is equivalent to \eqref{eq:s4B}.

Note that the key to solving problem (P7) lies in determining the azimuth angle $\phi_n$ in the feasible set defined by \eqref{eq:s5B} that is closest to $\phi_{\mathrm{T},n}$.
Let $\phi_{\mathrm{UT},n}\triangleq \arccos\left(\frac{\bar{\mathbf{q}}_{\mathrm{U},n}^T\bar{\mathbf{q}}_{\mathrm{T},n}}{\|\bar{\mathbf{q}}_{\mathrm{U},n}\|\|\bar{\mathbf{q}}_{\mathrm{T},n}\|}\right)\in [0,\pi]$ denote the azimuth-angle difference between the user direction $\vec{\mathbf{q}}_{\mathrm{U},n}$ and the target direction $\vec{\mathbf{q}}_{\mathrm{T},n}$, where $\bar{\mathbf{q}}_{\mathrm{U},n}\triangleq \vec{\mathbf{q}}_{\mathrm{U},n} - (\vec{\mathbf{q}}_{\mathrm{U},n}^T\mathbf{e}_3)\mathbf{e}_3$ and $\bar{\mathbf{q}}_{\mathrm{T},n}\triangleq \vec{\mathbf{q}}_{\mathrm{T},n} - (\vec{\mathbf{q}}_{\mathrm{T},n}^T\mathbf{e}_3)\mathbf{e}_3$ represent the projections of $\vec{\mathbf{q}}_{\mathrm{U},n}$ and $\vec{\mathbf{q}}_{\mathrm{T},n}$ on the $x$-$y$ plane, respectively.
If $\phi_{\mathrm{UT},n} \leq \Delta\phi_n$, then $\phi_n = \phi_{\mathrm{T},n}$ lies in the feasible set \eqref{eq:s5B}. Therefore, the optimal solution to problem (P7) is given by
\begin{align}
	\label{deqn_ex4g}
	\phi_n^{\star} = \phi_{\mathrm{T},n},\;\text{if}\; \phi_{\mathrm{UT},n} \leq \Delta\phi_n.
\end{align}
Otherwise, the optimal solution is attained at the boundary of the feasible set, i.e.,
\begin{align}
	\label{deqn_ex5g}
	\phi_n^{\star} = \operatorname{arg} \min_{\phi_n \in \{\phi_{n,1},\phi_{n,2}\}}{|\phi_{\mathrm{T},n} - \phi_n|},\;\text{if}\; \phi_{\mathrm{UT},n} > \Delta\phi_n,
\end{align}
where $\phi_{n,1}\triangleq \Delta\phi_n + \phi_{\mathrm{U},n} + 2k_{n,1}\pi$ with $k_{n,1} \in \{k|0\leq \Delta\phi_n + \phi_{\mathrm{U},n} + 2k\pi \leq 2\pi\}$, and $\phi_{n,2}\triangleq \phi_{\mathrm{U},n} - \Delta\phi_n + 2k_{n,2}\pi$ with $k_{n,2} \in \{k|0\leq \phi_{\mathrm{U},n} - \Delta\phi_n + 2k\pi \leq 2\pi\}$ are two solutions to the equation $|\phi_{\mathrm{U},n} - \phi_n| = \Delta\phi_n$.
Based on the above discussions, the maximum value of $\vec{\mathbf{f}}_n^T\vec{\mathbf{q}}_{\mathrm{T},n}$ is given by \eqref{deqn_ex6g}, shown at the top of this page, where $\psi_{\mathrm{T},n} \triangleq \arccos(\vec{\mathbf{q}}_{\mathrm{T},n}^T \mathbf{e}_3)$ denotes the angle between the target direction $\vec{\mathbf{q}}_{\mathrm{T},n}$ and the $z$-axis.
\begin{figure*}[ht]
	\begin{align}
		\label{deqn_ex6g}
		\max_{\vec{\mathbf{f}}_n\in\mathcal{F}_n}{\vec{\mathbf{f}}_n^T\vec{\mathbf{q}}_{\mathrm{T},n}} =
		\begin{cases}
			{\cos(\psi_{\mathrm{T},n} -\theta_{\mathrm{max}}),}&{\phi_{\mathrm{UT},n} \leq \Delta\phi_n} \\
			{\sin\theta_{\mathrm{max}}\sin\psi_{\mathrm{T},n}\cos(\phi_{\mathrm{UT},n}-\Delta\phi_n) + \cos\theta_{\mathrm{max}}\cos\psi_{\mathrm{T},n},}&{\phi_{\mathrm{UT},n} > \Delta\phi_n}.
		\end{cases}
	\end{align}\vspace{-0.3cm} \hrulefill
\end{figure*}

Unlike the maximum value of $\vec{\mathbf{f}}_n^T\vec{\mathbf{q}}_{\mathrm{T},n}$ derived in \eqref{deqn_ex2g} in Subsection~\ref{Planer_sector}, when the positions of both the user and the target are given (i.e., $\phi_{\mathrm{UT},n}$ and $\psi_{\mathrm{T},n}$ are fixed), the maximum value of $\vec{\mathbf{f}}_n^T\vec{\mathbf{q}}_{\mathrm{T},n}$ obtained in \eqref{deqn_ex6g} is determined not only by $R_{\mathrm{min}}$, through its influence on $\Delta\phi_n$ as shown in \eqref{deqn_ex3g}, but also by $\theta_{\mathrm{max}}$.
First, in the second case of \eqref{deqn_ex6g}, the maximum value of $\vec{\mathbf{f}}_n^T\vec{\mathbf{q}}_{\mathrm{T},n}$  increases with $\Delta \phi_n$, which corresponds to a smaller $R_{\mathrm{min}}$. Second, in both cases of \eqref{deqn_ex6g}, the maximum value of $\vec{\mathbf{f}}_n^T\vec{\mathbf{q}}_{\mathrm{T},n}$ first increases and then decreases as $\theta_{\mathrm{max}}$ increases. Specifically, for a given $R_{\mathrm{min}}$, this maximum is attained at $\theta_{\mathrm{max}} = \psi_{\mathrm{T},n}$ in the first case, and at $\theta_{\mathrm{max}} = \arctan \left(\frac{\sin\psi_{\mathrm{T},n}\cos(\phi_{\mathrm{UT},n}-\Delta\phi_n)}{\cos\psi_{\mathrm{T},n}}\right)$ in the second case.
This implies that the optimal $\vec{\mathbf{f}}_n$ does not lie on the boundary of the allowable rotation cone defined by \eqref{eq:s7C} when $\theta_{\mathrm{max}} \geq \psi_{\mathrm{T},n}$ for $\phi_{\mathrm{UT},n} \leq \Delta\phi_n$, and when $\theta_{\mathrm{max}} \geq \arctan \left(\frac{\sin\psi_{\mathrm{T},n}\cos(\phi_{\mathrm{UT},n}-\Delta\phi_n)}{\cos\psi_{\mathrm{T},n}}\right)$ for $\phi_{\mathrm{UT},n} > \Delta\phi_n$. Therefore, the optimal $\vec{\mathbf{f}}_n$ can only lie within the sector formed by $\vec{\mathbf{q}}_{\mathrm{U},n}$ and $\vec{\mathbf{q}}_{\mathrm{T},n}$, as discussed in Subsection~\ref{Planer_sector}. Hence, the optimal $\vec{\mathbf{f}}_n$ can be obtained from \eqref{deqn_ex1g}.

In summary, the above derivations show that the optimal RA pointing vector can be characterized by two geometric cases: it either lies within the sector spanned by the user and target directions, as given in \eqref{deqn_ex1g}, or on the boundary of the allowable rotation cone, as characterized by \eqref{eq:s4C} together with \eqref{deqn_ex4g} and \eqref{deqn_ex5g}. These two cases provide an intuitive geometric interpretation of the optimal RA orientation design.
The solution within the user-target sector corresponds to the case where the RA has sufficient rotational flexibility, so that its boresight can be adjusted within the sector spanned by the user direction and the target direction. In this case, the optimal pointing vector mainly reflects the trade-off between maintaining the required communication rate and enhancing the sensing echo power.
In contrast, when the desired pointing direction exceeds the allowable rotation range, the optimal pointing vector is forced to lie on the boundary of the allowable rotation cone. In this case, the RA cannot further rotate toward the desired target direction due to its limited rotational capability, and the optimal pointing vector is determined by the maximum allowable rotation range. Therefore, the two solution regions highlight two different governing factors of the optimal RA orientation: the communication-sensing trade-off and the hardware-imposed rotation limit, respectively.

\section{Multi-User and Regional Sensing Case}\label{Multi_user}
In this section, we consider a general RA-enabled ISAC system, where the RA-based BS simultaneously performs multi-user communication and regional sensing, i.e., $K > 1$ and the potential target is located within the sensing region~$\mathcal{A}$.
According to problem (P1), the RA pointing matrix $\mathbf{F}$, the transmit beamforming matrix $\mathbf{V}$, and the probing-signal covariance matrix $\mathbf{S}$ are jointly designed to maximize the minimum received echo signal power over the entire sensing region $\mathcal{A}$. In other words, the considered max-min design aims to enhance the worst-case sensing performance within the region and mitigate severe power imbalance across different spatial locations.
Thus, the objective function in \eqref{eq:0A} involves semi-infinite terms since the sensing region $\mathcal{A}$ is continuous, which makes solving the problem challenging.
To address this issue, we first discretize the continuous sensing region into $M$ spatial sampling points. Let $\mathbf{q}_{\mathrm{T},m}\in \mathbb{R}^{3\times 1}$ with $m\in \mathcal{M}\triangleq \{1,2,\dots,M\}$ represent the position of the $m$-th spatial sampling point.
In the following, we let $d_{\mathrm{U},k,n} \triangleq \|\mathbf{q}_{\mathrm{U},k} - \mathbf{w}_n\|$ and $\vec{\mathbf{q}}_{\mathrm{U},k,n} \triangleq \frac{\mathbf{q}_{\mathrm{U},k} - \mathbf{w}_n}{\|\mathbf{q}_{\mathrm{U},k} - \mathbf{w}_n\|}$ with $k\in \mathcal{K}$ denote the propagation distance and the normalized direction vector from RA~$n$ to user~$k$, respectively, and we let $d_{\mathrm{T},m,n} \triangleq \|\mathbf{q}_{\mathrm{T},m} - \mathbf{w}_n\|$ and $\vec{\mathbf{q}}_{\mathrm{T},m,n} \triangleq \frac{\mathbf{q}_{\mathrm{T},m} - \mathbf{w}_n}{\|\mathbf{q}_{\mathrm{T},m} - \mathbf{w}_n\|}$ with $m\in \mathcal{M}$ denote the propagation distance and the normalized direction vector from RA~$n$ to the $m$-th spatial sampling point, respectively.

By introducing the slack optimization variable $\eta$ to denote the minimum echo signal power within sensing region $\mathcal{A}$, problem (P1) can be approximated as
\begin{subequations}\label{eq:m1}
	\begin{alignat}{2}
		(\mathrm{P8}):\; \max_{\eta,\mathbf{F},\mathbf{V},\mathbf{S}\succeq \mathbf{0}} \quad & \eta & \label{eq:m1A}\\
		\mbox{s.t.} \quad
		& P(\mathbf{F},\mathbf{S},\mathbf{q}_{\mathrm{T},m}) \geq \eta,\; \forall m, \label{eq:m1B}\\
		& \eqref{eq:0B}-\eqref{eq:0F}. \label{eq:1C}
	\end{alignat}
\end{subequations}
Since constraints \eqref{eq:0B}, \eqref{eq:0F}, and \eqref{eq:m1B} are non-convex with respect to the optimization variables $\mathbf{F}$, $\mathbf{V}$, and $\mathbf{S}$, problem (P8) remains difficult to solve directly. To address this issue, we develop an AO algorithm to obtain an efficient suboptimal solution of problem (P8), based on the block coordinate descent (BCD) framework and the SCA technique. The main idea is to decompose problem (P8) into two subproblems. The first optimizes the beamforming and covariance matrices for a given RA pointing matrix. The second optimizes the RA pointing matrix for given beamforming and covariance matrices. These two subproblems are then solved iteratively, such that the variable blocks $\{\mathbf{V},\mathbf{S}\}$ and $\mathbf{F}$ are updated in an alternating manner until the objective value $\eta$ converges within a prescribed accuracy.
For notational convenience, in the remainder of this paper, we use $\mathbf{h}_{\mathrm{U},k}$ to denote the channel vector between the BS and user~$k$, i.e., $\mathbf{h}_{\mathrm{U}}(\mathbf{F},\mathbf{q}_{\mathrm{U},k})$, and $\mathbf{h}_{\mathrm{T},m}$ to denote the channel vector between the BS and the $m$-th spatial sampling point in sensing region~$\mathcal{A}$, i.e., $\mathbf{h}(\mathbf{F},\mathbf{q}_{\mathrm{T},m})$.

\subsection{Optimization of Beamforming and Covariance Matrices for Fixed RA Pointing Matrix}
In this subsection, we optimize the transmit beamforming matrix $\mathbf{V}$ and the probing-signal covariance matrix $\mathbf{S}$ with the RA pointing vectors of all RAs being fixed. Given the pointing matrix $\mathbf{F}$, problem (P8) reduces to
\begin{subequations}\label{eq:mv1}
	\begin{alignat}{2}
		(\mathrm{P9}):\;& \max_{\eta,\mathbf{V},\mathbf{S}\succeq \mathbf{0}} \; \eta & \label{eq:mv1A}\\
		\mbox{s.t.} \quad
		& |\beta_{\mathrm{T}}|^2 \|\mathbf{h}_{\mathrm{T},m}\|^2 \mathbf{h}_{\mathrm{T},m}^T \mathbf{S} \mathbf{h}_{\mathrm{T},m}^{\ast} \geq \eta, \; \forall m, \label{eq:mv1B}\\
		& |\mathbf{h}_{\mathrm{U},k}^T \mathbf{v}_k|^2 \geq \bar{\gamma}_k\hspace{-0.1cm}\left(\sum_{j\ne k}{|\mathbf{h}_{\mathrm{U},k}^T \mathbf{v}_j|^2} + \sigma_k^2\right)\hspace{-0.1cm}, \;\forall k,\hspace{-0.1cm}\label{eq:mv1C}\\
		& \eqref{eq:0C}, \eqref{eq:0D}, \label{eq:mv1D}
	\end{alignat}
\end{subequations}
where $\bar{\gamma}_k \triangleq 2^{R_{\mathrm{min},k}} - 1$ denotes the required minimum SNR of user $k$.
SDR can be adopted to overcome non-convex constraint \eqref{eq:mv1C}. First, we define an auxiliary variable $\mathbf{V}_k \triangleq \mathbf{v}_k\mathbf{v}_k^H$, which satisfies $\mathbf{V}_k \succeq \mathbf{0}$ and $\operatorname{rank}(\mathbf{V}_k) = 1$. Then, by relaxing the non-convex rank-one constraint, problem (P9) can be approximated by
\begin{subequations}\label{eq:mv2}
	\begin{alignat}{2}
		\hspace{-0.5
			cm}(\mathrm{P10}):&\; \max_{\eta,\mathbf{S},\{\mathbf{V}_k\}} \; \eta & \label{eq:mv2A}\\
		\mbox{s.t.} \quad
		& \frac{1}{\bar{\gamma}_k} \mathbf{h}_{\mathrm{U},k}^T \mathbf{V}_k \mathbf{h}^{\ast}_{\mathrm{U},k} \geq \sum_{j\ne k}{\mathbf{h}_{\mathrm{U},k}^T \mathbf{V}_j \mathbf{h}^{\ast}_{\mathrm{U},k}} + \sigma_k^2,\; \forall k,\hspace{-0.15cm}\label{eq:mv2B}\\
		& \sum_{k=1}^{K}{\operatorname{Tr}(\mathbf{V}_k)} \leq P_{\mathrm{max,c}}, \label{eq:mv2C}\\
		& \mathbf{S}\succeq \mathbf{0},\; \mathbf{V}_k\succeq \mathbf{0}, \; \forall k,\label{eq:mv2D}\\
		& \eqref{eq:0D}, \eqref{eq:mv1B}. \label{eq:mv2E}
	\end{alignat}
\end{subequations}
Note that problem (P10) is a standard semidefinite program (SDP), which can be optimally solved using the CVX solver~\cite{Boyd2004Convex}.

Although problem (P10) may not yield a rank-one solution to $\mathbf{V}_k$, we can always construct the following rank-one solution that achieves the same objective value~\cite{Liu2020Joint}:
\begin{align}
	\label{deqn_ex1e}
	\mathbf{v}_k^{\star} = \left(\mathbf{h}_{\mathrm{U},k}^T \tilde{\mathbf{V}}_k \mathbf{h}^{\ast}_{\mathrm{U},k}\right)^{-\frac{1}{2}} \tilde{\mathbf{V}}_k \mathbf{h}^{\ast}_{\mathrm{U},k},\; \forall k,
\end{align}
where $\tilde{\mathbf{V}}_k$ denotes the globally optimal $\mathbf{V}_k$ obtained by solving problem (P10). Thus, the rank-one solution to $\mathbf{V}_k$ is given by $\mathbf{V}_k^{\star} = \mathbf{v}_k^{\star} (\mathbf{v}_k^{\star})^H$.

\subsection{Optimization of RA Pointing Matrix for Fixed Beamforming and Covariance Matrices}
In this subsection, we optimize the pointing matrix $\mathbf{F}$ with the transmit beamforming and probing-signal covariance matrix being fixed. Given transmit beamforming matrix $\mathbf{V}$ and probing-signal covariance matrix $\mathbf{S}$, problem (P8) reduces to
\begin{subequations}\label{eq:mf1}
	\begin{alignat}{2}
		(\mathrm{P11}):\; \max_{\eta,\mathbf{F}} \quad& \eta & \label{eq:mf1A}\\
		\mbox{s.t.} \quad
		& |\beta_{\mathrm{T}}|^2 \mathbf{h}_{\mathrm{T},m}^T \mathbf{S} \mathbf{h}_{\mathrm{T},m}^{\ast} \geq \frac{\eta}{\|\mathbf{h}_{\mathrm{T},m}\|^2},\; \forall m, \label{eq:mf1B}\\
		& \cos\theta_{\mathrm{max}} \leq \vec{\mathbf{f}}_n^T \mathbf{e}_3 \leq 1,\; \forall n, \label{eq:mf1D}\\
		& \eqref{eq:0F}, \eqref{eq:mv2B}, \label{eq:mf1E}
	\end{alignat}
\end{subequations}
where constraint \eqref{eq:mf1D} is equivalent to \eqref{eq:0E}. Since constraints \eqref{eq:mf1B} and \eqref{eq:mv2B} are non-convex with respect to $\mathbf{F}$, and the equality constraint \eqref{eq:0F} is not affine, the above subproblem is challenging to solve directly. Specifically, the main challenge in solving problem (P11) lies in handling the Hermitian quadratic forms present in constraints \eqref{eq:mf1B} and \eqref{eq:mv2B}.

First, let $\mathbf{a}_{\mathrm{U},k}\in \mathbb{R}^{N\times 1}$ and $\mathbf{p}_{\mathrm{U},k}\in \mathbb{C}^{N\times 1}$ denote the amplitude and phase-shift components of communication channel vector $\mathbf{h}_{\mathrm{U},k}$, respectively. Therefore, $\mathbf{h}_{\mathrm{U},k}$ can be expressed as
\begin{align}
	\label{deqn_ex1f}
	\mathbf{h}_{\mathrm{U},k} = \mathbf{a}_{\mathrm{U},k} \odot \mathbf{p}_{\mathrm{U},k} \overset{(a)}{=} \operatorname{diag}\left(\mathbf{p}_{\mathrm{U},k}\right) \mathbf{a}_{\mathrm{U},k},
\end{align}
where the $n$-th entries of $\mathbf{a}_{\mathrm{U},k}$ and $\mathbf{p}_{\mathrm{U},k}$ are given by $a_{\mathrm{U},k,n} = \beta_{\mathrm{U},k,n}[\vec{\mathbf{f}}_n^T \vec{\mathbf{q}}_{\mathrm{U},k,n}]_{+}^{\rho}$ and $p_{\mathrm{U},k,n} = e^{-j\frac{2\pi}{\lambda}d_{\mathrm{U},k,n}}$, respectively, with $\beta_{\mathrm{U},k,n} = \frac{\sqrt{\beta_0 G_{\mathrm{max}}}}{d_{\mathrm{U},k,n}}$ denoting the channel coefficient of the link between user~$k$ and RA~$n$.
In addition, equality $(a)$ holds since $\operatorname{diag}(\mathbf{x})\cdot \mathbf{y} = \mathbf{x} \odot \mathbf{y}$ for column vectors $\mathbf{x}$ and $\mathbf{y}$ of equal size.
Similarly, by denoting the amplitude and phase-shift components of sensing channel vector $\mathbf{h}_{\mathrm{T},m}$ as $\mathbf{a}_{\mathrm{T},m} = [a_{\mathrm{T},m,1},a_{\mathrm{T},m,2},\dots,a_{\mathrm{T},m,N}]^T\in \mathbb{R}^{N\times 1}$ and $\mathbf{p}_{\mathrm{T},m} = [p_{\mathrm{T},m,1},p_{\mathrm{T},m,2},\dots,p_{\mathrm{T},m,N}]^T\in \mathbb{C}^{N\times 1}$, respectively, with $a_{\mathrm{T},m,n} = \beta_{\mathrm{T},m,n}[\vec{\mathbf{f}}_n^T \vec{\mathbf{q}}_{\mathrm{T},m,n}]_{+}^{\rho}$ and $p_{\mathrm{T},m,n} = e^{-j\frac{2\pi}{\lambda}d_{\mathrm{T},m,n}}$, where $\beta_{\mathrm{T},m,n} = \frac{\sqrt{\beta_0 G_{\mathrm{max}}}}{d_{\mathrm{T},m,n}}$ is the channel coefficient of the link between the $m$-th spatial sampling point within sensing region $\mathcal{A}$ and RA~$n$, $\mathbf{h}_{\mathrm{T},m}$ can be written as
\begin{align}
	\label{deqn_ex0f}
	\mathbf{h}_{\mathrm{T},m} = \operatorname{diag}\left(\mathbf{p}_{\mathrm{T},m}\right) \mathbf{a}_{\mathrm{T},m}.
\end{align}
Then, to tackle the non-convexity of problem (P11), we introduce the following proposition.

\textit{Proposition 1:} For $k,j\in \mathcal{K}$, Hermitian quadratic form $\mathbf{h}_{\mathrm{U},k}^T \mathbf{V}_j \mathbf{h}^{\ast}_{\mathrm{U},k}$ in \eqref{eq:mv2B} is convex with respect to $\mathbf{a}_{\mathrm{U},k}$.

\begin{IEEEproof}
	According to \eqref{deqn_ex1f} and \eqref{deqn_ex0f}, $\mathbf{h}_{\mathrm{U},k}^T \mathbf{V}_j \mathbf{h}^{\ast}_{\mathrm{U},k}$ can be written as
\begin{align}
	\label{deqn_ex2f}
	\mathbf{h}_{\mathrm{U},k}^T \mathbf{V}_j \mathbf{h}^{\ast}_{\mathrm{U},k} =& \mathbf{a}_{\mathrm{U},k}^T \operatorname{diag}\left( \mathbf{p}_{\mathrm{U},k} \right) \mathbf{V}_j \operatorname{diag}\left(\mathbf{p}^{\ast}_{\mathrm{U},k} \right) \mathbf{a}^{\ast}_{\mathrm{U},k} \nonumber\\
	\overset{(b)}{=} & \mathbf{a}_{\mathrm{U},k}^T \underbrace{\mathbf{V}_j \odot \left(\mathbf{p}_{\mathrm{U},k} \mathbf{p}_{\mathrm{U},k}^H \right)}_{\bar{\mathbf{V}}_{k,j}} \mathbf{a}_{\mathrm{U},k} \nonumber\\
	= & \mathbf{a}_{\mathrm{U},k}^T \bar{\mathbf{V}}_{k,j} \mathbf{a}_{\mathrm{U},k},
\end{align}
where equality $(b)$ holds since $\operatorname{diag}(\mathbf{x})\mathbf{U}\operatorname{diag}(\mathbf{y}) = \mathbf{U} \odot \left(\mathbf{x} \mathbf{y}^T \right)$ for $\mathbf{x},\mathbf{y}\in \mathbb{C}^{N\times 1}$ and $\mathbf{U}\in \mathbb{C}^{N\times N}$~\cite{Horn1985Matrix}. Since the Hadamard product of two positive semi-definite matrices is also positive semi-definite, it follows that $\bar{\mathbf{V}}_{k,j}\triangleq \mathbf{V}_j \odot (\mathbf{p}_{\mathrm{U},k} \mathbf{p}_{\mathrm{U},k}^H)$ is a positive semi-definite matrix, where $\mathbf{V}_j = \mathbf{v}_j\mathbf{v}_j^H \succeq \mathbf{0}$ and $\mathbf{p}_{\mathrm{U},k} \mathbf{p}_{\mathrm{U},k}^H \succeq \mathbf{0}$. Therefore, $\mathbf{a}_{\mathrm{U},k}^T \bar{\mathbf{V}}_{k,j} \mathbf{a}_{\mathrm{U},k}$ in \eqref{deqn_ex2f} is convex with respect to $\mathbf{a}_{\mathrm{U},k}$.
\end{IEEEproof}

Similarly, by defining $\bar{\mathbf{S}}_m\triangleq \mathbf{S}\odot (\mathbf{p}_{\mathrm{T},m} \mathbf{p}_{\mathrm{T},m}^H)$, Hermitian quadratic form $\mathbf{h}_{\mathrm{T},m}^T \mathbf{S} \mathbf{h}_{\mathrm{T},m}^{\ast}$ in \eqref{eq:mf1B} can be written as
\begin{align}
	\label{deqn_ex3f}
	\mathbf{h}_{\mathrm{T},m}^T \mathbf{S} \mathbf{h}_{\mathrm{T},m}^{\ast} = \mathbf{a}_{\mathrm{T},m}^T \bar{\mathbf{S}}_{m} \mathbf{a}_{\mathrm{T},m},\; \forall m.
\end{align}
We can also prove that $\bar{\mathbf{S}}_m$ is a positive semi-definite matrix, and thus, $\mathbf{h}_{\mathrm{T},m}^T \mathbf{S} \mathbf{h}_{\mathrm{T},m}^{\ast}$ is convex with respect to $\mathbf{a}_{\mathrm{T},m}$.

To transform problem (P11) into a more tractable form, we introduce auxiliary variables $\tilde{\eta}$, $\left\{\tilde{\mathbf{a}}_{\mathrm{U},k}\in \mathbb{R}^{N\times 1},\; \forall k\in \mathcal{K} \right\}$, $\left\{\tilde{\mathbf{a}}_{\mathrm{T},m}\in \mathbb{R}^{N\times 1},\; \forall m\in \mathcal{M} \right\}$, and $\left\{\delta_m,\; \forall m\in \mathcal{M}\right\}$, and reformulate problem (P11) as follows.
\vspace{-0.1cm}
\begin{subequations}\label{eq:mf2}
	\begin{alignat}{2}
		\hspace{-0.25cm}(\mathrm{P12}):\;& \max_{\Xi} \; \tilde{\eta} & \label{eq:mf2A}\\
		\mbox{s.t.} \quad
		& |\beta_{\mathrm{T}}|^2\tilde{\mathbf{a}}_{\mathrm{T},m}^T \bar{\mathbf{S}}_m \tilde{\mathbf{a}}_{\mathrm{T},m} \geq \frac{\tilde{\eta}^2}{\delta_m},\; \forall m, \label{eq:mf2B}\\
		& \frac{1}{\bar{\gamma}_k} \tilde{\mathbf{a}}_{\mathrm{U},k}^T \bar{\mathbf{V}}_{k,k} \tilde{\mathbf{a}}_{\mathrm{U},k} \hspace{-0.05cm}\geq\hspace{-0.05cm} \sum_{j\ne k}{\tilde{\mathbf{a}}_{\mathrm{U},k}^T \bar{\mathbf{V}}_{k,j} \tilde{\mathbf{a}}_{\mathrm{U},k}} \hspace{-0.05cm}+\hspace{-0.05cm} \sigma_k^2,\; \forall k,\hspace{-0.2cm} \label{eq:mf2C}\\
		& \|\tilde{\mathbf{a}}_{\mathrm{T},m}\|^2 \geq \delta_m,\; \forall m,\label{eq:mf2D}\\	
		& \tilde{a}_{\mathrm{T},m,n} = a_{\mathrm{T},m,n}, \; \forall m,n, \label{eq:mf2E}\\
		& \tilde{a}_{\mathrm{U},k,n} = a_{\mathrm{U},k,n}, \; \forall k,n, \label{eq:mf2F}\\
		& \eqref{eq:0F}, \eqref{eq:mf1D}, \label{eq:mf2G}
	\end{alignat}
\end{subequations}
where we exploited \eqref{deqn_ex2f} and \eqref{deqn_ex3f}, and defined $\Xi\triangleq \left\{\tilde{\eta},\mathbf{F},\tilde{\mathbf{a}}_{\mathrm{U},k},\tilde{\mathbf{a}}_{\mathrm{T},m},\delta_m\right\}$ as the set of all optimization variables.
To facilitate the convex reformulation, by letting $\tilde{\eta}^2 = \eta$, the right-hand side of \eqref{eq:mf1B} was equivalently rewritten into a convex quadratic-over-linear form in \eqref{eq:mf2B}. Since $\tilde{\eta}^2$ is monotonically increasing with respect to $\tilde{\eta}$ for $\tilde{\eta}\geq 0$, maximizing $\tilde{\eta}$ in (P12) is equivalent to maximizing the original variable $\eta$ in (P11).
In \eqref{eq:mf2E} and \eqref{eq:mf2F}, $\tilde{a}_{\mathrm{T},m,n}$ and $\tilde{a}_{\mathrm{U},k,n}$ denote the $n$-th entries of $\tilde{\mathbf{a}}_{\mathrm{T},m}$ and $\tilde{\mathbf{a}}_{\mathrm{U},k}$, respectively.
Additionally, we can prove that problem (P12) is equivalent to problem (P11) by contradiction.
Problem (P12) is still non-convex due to the non-convexity of constraints \eqref{eq:0F} and \eqref{eq:mf2B}--\eqref{eq:mf2F}.
Therefore, in the following, we suitably transform and approximate these non-convex constraints to facilitate tractable optimization.

Since the Hermitian quadratic forms $\tilde{\mathbf{a}}_{\mathrm{T},m}^T \bar{\mathbf{S}}_m \tilde{\mathbf{a}}_{\mathrm{T},m}$ and $\tilde{\mathbf{a}}_{\mathrm{U},k}^T \bar{\mathbf{V}}_{k,k} \tilde{\mathbf{a}}_{\mathrm{U},k}$ on the left-hand side of \eqref{eq:mf2B} and \eqref{eq:mf2C} are convex with respect to $\tilde{\mathbf{a}}_{\mathrm{T},m}$ and $\tilde{\mathbf{a}}_{\mathrm{U},k}$, respectively, we can tackle the non-convexity of constraints \eqref{eq:mf2B} and \eqref{eq:mf2C} by applying SCA. This allows us to replace the non-convex constraints with suitable convex approximations and then solve the approximated problem iteratively until its objective function converges.
As the first-order Taylor expansion of a convex function is a global under-estimator, if we define $\tilde{\mathbf{a}}_{\mathrm{T},m}^{(i)}$ and $\tilde{\mathbf{a}}_{\mathrm{U},k}^{(i)}$ as the solutions of $\tilde{\mathbf{a}}_{\mathrm{T},m}$ and $\tilde{\mathbf{a}}_{\mathrm{U},k}$ in the $i$-th iteration, $\tilde{\mathbf{a}}_{\mathrm{T},m}^T \bar{\mathbf{S}}_m \tilde{\mathbf{a}}_{\mathrm{T},m}$ and $\tilde{\mathbf{a}}_{\mathrm{U},k}^T \bar{\mathbf{V}}_{k,k} \tilde{\mathbf{a}}_{\mathrm{U},k}$ can be lower-bounded using their first-order Taylor expansions at $\tilde{\mathbf{a}}_{\mathrm{T},m}^{(i)}$ and $\tilde{\mathbf{a}}_{\mathrm{U},k}^{(i)}$ as $\Gamma_m^{(i+1)}\left(\tilde{\mathbf{a}}_{\mathrm{T},m}\right)$ and $\Lambda_k^{(i+1)}\left(\tilde{\mathbf{a}}_{\mathrm{U},k}\right)$ given in \eqref{deqn_ex4f} and \eqref{deqn_ex5f}, respectively, shown at the top of this page.
\begin{figure*}[ht]
\begin{align}
	\hspace{-0.45cm}\tilde{\mathbf{a}}_{\mathrm{T},m}^T \bar{\mathbf{S}}_m \tilde{\mathbf{a}}_{\mathrm{T},m} \hspace{-0.05cm}\geq& (\tilde{\mathbf{a}}_{\mathrm{T},m}^{(i)})^T \bar{\mathbf{S}}_m \tilde{\mathbf{a}}_{\mathrm{T},m}^{(i)} \hspace{-0.05cm}+\hspace{-0.05cm} 2(\tilde{\mathbf{a}}_{\mathrm{T},m}^{(i)})^T \bar{\mathbf{S}}_m \hspace{-0.1cm}\left(\tilde{\mathbf{a}}_{\mathrm{T},m} \hspace{-0.05cm}-\hspace{-0.05cm} \tilde{\mathbf{a}}_{\mathrm{T},m}^{(i)}\right) \hspace{-0.05cm}=\hspace{-0.05cm} 2(\tilde{\mathbf{a}}_{\mathrm{T},m}^{(i)})^T \bar{\mathbf{S}}_m \tilde{\mathbf{a}}_{\mathrm{T},m} \hspace{-0.05cm}-\hspace{-0.05cm} (\tilde{\mathbf{a}}_{\mathrm{T},m}^{(i)})^T \bar{\mathbf{S}}_m \tilde{\mathbf{a}}_{\mathrm{T},m}^{(i)} \hspace{-0.05cm}\triangleq\hspace{-0.05cm} \Gamma_m^{(i+1)}\hspace{-0.05cm}\left(\tilde{\mathbf{a}}_{\mathrm{T},m}\right)\hspace{-0.05cm},\hspace{-0.2cm}\label{deqn_ex4f}\\
	\hspace{-0.45cm}\tilde{\mathbf{a}}_{\mathrm{U},k}^T \bar{\mathbf{V}}_{k,k} \tilde{\mathbf{a}}_{\mathrm{U},k} \hspace{-0.05cm}\geq& (\tilde{\mathbf{a}}_{\mathrm{U},k}^{(i)})^T \bar{\mathbf{V}}_{k,k} \tilde{\mathbf{a}}_{\mathrm{U},k}^{(i)} \hspace{-0.05cm}+\hspace{-0.05cm} 2(\tilde{\mathbf{a}}_{\mathrm{U},k}^{(i)})^T \bar{\mathbf{V}}_{k,k} \hspace{-0.1cm}\left(\tilde{\mathbf{a}}_{\mathrm{U},k} \hspace{-0.05cm}-\hspace{-0.05cm} \tilde{\mathbf{a}}_{\mathrm{U},k}^{(i)}\right) \hspace{-0.05cm}=\hspace{-0.05cm} 2(\tilde{\mathbf{a}}_{\mathrm{U},k}^{(i)})^T \bar{\mathbf{V}}_{k,k} \tilde{\mathbf{a}}_{\mathrm{U},k} \hspace{-0.05cm}-\hspace{-0.05cm} (\tilde{\mathbf{a}}_{\mathrm{U},k}^{(i)})^T \bar{\mathbf{V}}_{k,k} \tilde{\mathbf{a}}_{\mathrm{U},k}^{(i)}\hspace{-0.05cm}\triangleq\hspace{-0.05cm} \Lambda_k^{(i+1)}\hspace{-0.05cm}\left(\tilde{\mathbf{a}}_{\mathrm{U},k}\right)\hspace{-0.05cm}.\hspace{-0.2cm}\label{deqn_ex5f}
\end{align} \vspace{-0.3cm} \hrulefill
\end{figure*}
Note that, in \eqref{deqn_ex4f} and \eqref{deqn_ex5f}, equality holds if and only if $\tilde{\mathbf{a}}_{\mathrm{T},m} = \tilde{\mathbf{a}}_{\mathrm{T},m}^{(i)}$ and $\tilde{\mathbf{a}}_{\mathrm{U},k} = \tilde{\mathbf{a}}_{\mathrm{U},k}^{(i)}$, respectively. Accordingly, in the $(i+1)$-th SCA iteration, constraints \eqref{eq:mf2B} and \eqref{eq:mf2C} are approximated by the following two convex quadratic constraints:
\begin{align}
	|\beta_{\mathrm{T}}|^2\Gamma_m^{(i+1)}\left(\tilde{\mathbf{a}}_{\mathrm{T},m}\right) &\geq \frac{\tilde{\eta}^2}{\delta_m},\; \forall m, \label{deqn_ex6f}\\		
	\frac{1}{\bar{\gamma}_k}\Lambda_k^{(i+1)}\left(\tilde{\mathbf{a}}_{\mathrm{U},k}\right) &\geq \sum_{j\ne k}{\tilde{\mathbf{a}}_{\mathrm{U},k}^T \bar{\mathbf{V}}_{k,j} \tilde{\mathbf{a}}_{\mathrm{U},k}} + \sigma_k^2,\; \forall k.\label{deqn_ex7f}
\end{align}
Similarly, since $\|\tilde{\mathbf{a}}_{\mathrm{T},m}\|^2$ is convex with respect to $\tilde{\mathbf{a}}_{\mathrm{T},m}$, constraint \eqref{eq:mf2D} can be approximated by
\begin{align}
	\label{deqn_ex8f}
	2(\tilde{\mathbf{a}}_{\mathrm{T},m}^{(i)})^T\tilde{\mathbf{a}}_{\mathrm{T},m} - \|\tilde{\mathbf{a}}_{\mathrm{T},m}^{(i)}\|^2 \geq \delta_m,\; \forall m,
\end{align}
where, in \eqref{deqn_ex8f}, equality holds if and only if $\tilde{\mathbf{a}}_{\mathrm{T},m} = \tilde{\mathbf{a}}_{\mathrm{T},m}^{(i)}$.

Additionally, by using the first-order Taylor expansions of the non-affine terms on the right-hand sides of \eqref{eq:mf2E} and \eqref{eq:mf2F} at $\vec{\mathbf{f}}_n^{(i)}$, which denotes the solution of $\vec{\mathbf{f}}_n$ obtained in the $i$-th iteration, constraints \eqref{eq:mf2E} and \eqref{eq:mf2F} can be linearized as
\begin{align}
	\tilde{a}_{\mathrm{T},m,n} &\approx a_{\mathrm{T},m,n}^{(i)} + (\dot{\mathbf{a}}_{\mathrm{T},m,n}^{(i)})^T(\vec{\mathbf{f}}_n - \vec{\mathbf{f}}_n^{(i)}),\; \forall m,n, \label{deqn_ex9f}\\
	\tilde{a}_{\mathrm{U},k,n} &\approx a_{\mathrm{U},k,n}^{(i)} + (\dot{\mathbf{a}}_{\mathrm{U},k,n}^{(i)})^T (\vec{\mathbf{f}}_n - \vec{\mathbf{f}}_n^{(i)}),\; \forall k,n, \label{deqn_ex10f}
\end{align}
respectively, where $a_{\mathrm{T},m,n}^{(i)} = \beta_{\mathrm{T},m,n} [ (\vec{\mathbf{f}}_n^{(i)} )^T \vec{\mathbf{q}}_{\mathrm{T},m,n}]_{+}^{\rho}$ and $a_{\mathrm{U},k,n}^{(i)} = \beta_{\mathrm{U},k,n} [ (\vec{\mathbf{f}}_n^{(i)} )^T \vec{\mathbf{q}}_{\mathrm{U},k,n}]_{+}^{\rho}$ are the values of $a_{\mathrm{T},m,n}$ and $a_{\mathrm{U},k,n}$ obtained in the $i$-th iteration, respectively, and $\dot{\mathbf{a}}_{\mathrm{T},m,n}^{(i)} = \beta_{\mathrm{T},m,n} \rho[(\vec{\mathbf{f}}_n^{(i)})^T \vec{\mathbf{q}}_{\mathrm{T},m,n}]_{+}^{\rho-1}\vec{\mathbf{q}}_{\mathrm{T},m,n}$ and $\dot{\mathbf{a}}_{\mathrm{U},k,n}^{(i)} = \beta_{\mathrm{U},k,n} \rho[(\vec{\mathbf{f}}_n^{(i)})^T \vec{\mathbf{q}}_{\mathrm{U},k,n}]_{+}^{\rho-1}\vec{\mathbf{q}}_{\mathrm{U},k,n}$ denote the first-order derivatives of $a_{\mathrm{T},m,n}^{(i)}$ and $a_{\mathrm{U},k,n}^{(i)}$ relative to $\vec{\mathbf{f}}_n^{(i)}$, respectively.

Furthermore, for equality constraint \eqref{eq:0F}, by introducing two sufficiently small positive constants $\zeta \geq 0$ and $\xi \geq 0$, it can be transformed into the following two inequality constraints:
\begin{subequations}\label{deqn_ex12f}
	\begin{align}
		\|\vec{\mathbf{f}}_n\|^2 &\leq 1 + \zeta,\;\forall n, \label{deqn_ex12f1}\\
		\|\vec{\mathbf{f}}_n\|^2 &\geq 1 - \xi,\;\forall n. \label{deqn_ex12f2}
	\end{align}
\end{subequations}
We note that \eqref{deqn_ex12f} is equivalent to equality constraint \eqref{eq:0F} when $\zeta \to 0$ and $\xi \to 0$. Similar to \eqref{deqn_ex8f}, non-convex constraint \eqref{deqn_ex12f2} can be approximated by
\begin{align}
	\label{deqn_ex13f}
	2(\vec{\mathbf{f}}_n^{(i)})^T \vec{\mathbf{f}}_n - \|\vec{\mathbf{f}}_n^{(i)}\|^2 \geq 1 - \xi,\;\forall n,
\end{align}
where equality holds if and only if $\vec{\mathbf{f}}_n = \vec{\mathbf{f}}_n^{(i)}$.


According to the above derivation, by replacing constraints \eqref{eq:mf2B}--\eqref{eq:mf2F} with their convex approximations given in \eqref{deqn_ex6f}--\eqref{deqn_ex10f}, and relaxing equality constraint \eqref{eq:0F} according to \eqref{deqn_ex12f1} and \eqref{deqn_ex13f}, the approximated problem in the $(i+1)$-th iteration can be constructed as
\begin{subequations}\label{eq:mf4}
	\begin{alignat}{2}
		(\mathrm{P13}):\; \max_{\Xi} \quad& \tilde{\eta} \label{eq:mf4A}\\
		\mbox{s.t.} \quad
		& \eqref{eq:mf1D}, \eqref{deqn_ex6f}\hspace{-0.08cm}-\hspace{-0.08cm}\eqref{deqn_ex10f}, \eqref{deqn_ex12f1}, \eqref{deqn_ex13f}. \label{eq:mf4B}
	\end{alignat}
\end{subequations}
It can be verified that problem (P13) is a convex optimization problem, which can be solved by the CVX solver~\cite{Boyd2004Convex}. Since problem (P13) may not obtain a unit-norm pointing vector due to the relaxation of equality constraint \eqref{eq:0F}, we apply the normalization $\vec{\mathbf{f}}_n^{\star} = {\tilde{\mathbf{f}}_n}/{\|\tilde{\mathbf{f}}_n\|},\; \forall n\in \mathcal{N}$, where $\tilde{\mathbf{f}}_n$ is the solution of $\vec{\mathbf{f}}_n$ obtained by solving problem (P13). Additionally, after solving problem (P13), the corresponding objective value of the original subproblem, i.e., problem (P11), is recovered as $\eta^{\star} = \tilde{\eta}^2$.

\subsection{Overall Algorithm}
Based on the solutions of the two subproblems presented above, we propose an AO algorithm for solving problem (P8) based on the BCD method. The details are summarized in Algorithm~\ref{alg1}.
In each iteration, problem (P10) is optimally solved, while the SCA approximations in problem (P13) are constructed at the current feasible point. Thus, the objective value $\eta$ is non-decreasing over the iterations. Moreover, $\eta$ is upper-bounded under the finite communication and sensing transmit-power constraints, and thus the objective value sequence is guaranteed to converge. Since problem (P8) is non-convex, global optimality cannot be guaranteed. Under the standard SCA conditions, Algorithm~\ref{alg1} converges to a stationary point of problem (P8).
Subproblems (P10) and (P13) are solved by the CVX solver with complexity orders of $\mathcal{O}\left((3\bar{N})^{3.5}\right)$ and $\mathcal{O}\left(\bar{N}^{3.5}\operatorname{ln}(1/\varepsilon_1)\right)$, respectively~\cite{Boyd2004Convex,Bubeck2015Convex}, where $\bar{N}$ and $\varepsilon_1$ denote the number of optimization variables and the prescribed accuracy tolerance of the interior-point method, respectively. Therefore, the complexity order of Algorithm~\ref{alg1} is $\mathcal{O}\left(L ((3N(K+1))^{3.5} \hspace{-0.05cm}+\hspace{-0.05cm} ((K+M+1)N)^{3.5}\operatorname{ln}(1/\varepsilon_1)) \right)$, where $L$ denotes the number of iterations required for the algorithm to converge.
\begin{algorithm}[t!]
	\caption{Proposed AO Algorithm for Solving (P8).}
	\begin{algorithmic}[1] \label{alg1}
		\STATE Input: Pointing matrix $\mathbf{F}^{(0)}$, minimum echo signal power $\eta^{(0)}$, threshold $\varepsilon_2 > 0$, and maximum iteration number~$L$.
		\STATE Initialization: $i\gets 0$.
		\REPEAT
		\STATE Given $\mathbf{F}^{(i)}$, obtain $\mathbf{V}^{(i+1)}$ and $\mathbf{S}^{(i+1)}$ by solving (P10).
		\STATE Construct the rank-one solution to $\mathbf{V}^{(i+1)}$ with \eqref{deqn_ex1e}.
		\STATE Given $\mathbf{V}^{(i+1)}$, $\mathbf{S}^{(i+1)}$, and $\mathbf{F}^{(i)}$, obtain $\mathbf{F}^{(i+1)}$ and $\eta^{(i+1)}$ by solving (P13).
		\STATE Normalize the pointing vectors: $\vec{\mathbf{f}}_n^{(i+1)} = \frac{\vec{\mathbf{f}}_n^{(i+1)}}{\|\vec{\mathbf{f}}_n^{(i+1)}\|},\; \forall n$.
		\STATE Update $i=i+1$.
		\UNTIL $|\frac{\eta^{(i)} - \eta^{(i-1)}}{\eta^{(i-1)}}| \leq \varepsilon_2$ or $i>L$.
		\STATE Output: $\eta = \eta^{(i)}$, $\mathbf{V} = \mathbf{V}^{(i)}$, $\mathbf{S} = \mathbf{S}^{(i)}$, and $\mathbf{F} = \mathbf{F}^{(i)}$.
	\end{algorithmic}
\end{algorithm}

\section{Simulation Results}\label{Simulation}
In this section, we present simulation results to evaluate the performance of the proposed RA-enabled ISAC system and the proposed optimization algorithm. For our simulations, we consider an ISAC system with $K$ communication users and a square sensing region of size $r\times r\; \text{m}^2$, where $r$ denotes the side length of the sensing region.
The system operates at 2.4 GHz with a wavelength of $\lambda = 0.125$~m. The BS is equipped with a uniform planar array (UPA) consisting of $N = 4\times 4$ RAs, with an inter-element spacing of $\frac{\lambda}{2}$. In addition, the UPA is deployed on the $x$-$y$ plane of a Cartesian coordinate system and centered at the origin.
The noise powers at the BS and all users are assumed to be identical, i.e., $\sigma_{\mathrm{s}}^2 = \sigma_k^2 = -80$~dBm, $\forall k\in \mathcal{K}$. The maximum average transmit powers for the communication and probing signals are set to $P_{\mathrm{max,c}} = 20$~dBm and $P_{\mathrm{max,s}} = 40$~dBm, respectively.
Unless otherwise stated, the maximum allowable rotation angle of each RA is set to $\theta_{\mathrm{max}} = \frac{\pi}{6}$, the antenna directivity factor is set to $\rho = 4$, and the side length of the sensing region is set to $r = 6$~m. The required minimum communication rates of all users are given by $R_{\mathrm{min},k} = R_{\mathrm{min}}$, $\forall k\in \mathcal{K}$, where $R_{\mathrm{min}}$ may vary across different simulation settings.
Additionally, the proposed AO algorithm is initialized with the reference RA orientation, where all RAs point toward the default boresight direction, i.e., $\vec{\mathbf{f}}_n^{(0)} = \mathbf{e}_3,\; \forall n$. This deterministic initialization directly satisfies the RA rotation constraints and avoids introducing randomness into the iterative optimization.
All simulations are performed using MATLAB R2024a and a computer equipped with an Intel Core i5-14400F CPU and 32~GB of RAM. The convex optimization subproblems are solved using CVX with the SDPT3 solver.

\subsection{Single-User and Point-Target Case}
\begin{figure}[!t]  \centering
	\includegraphics[width=2.6in]{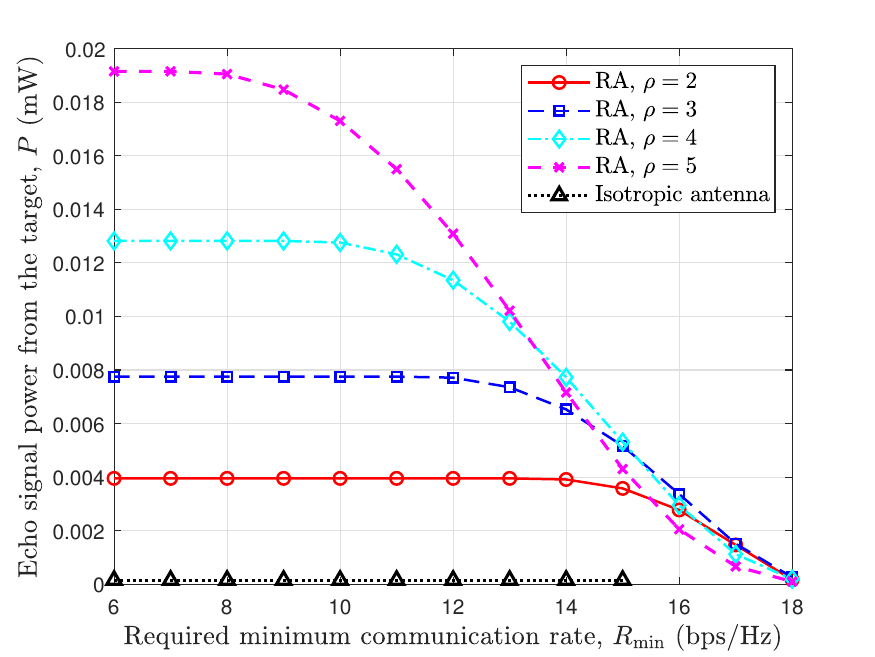}\vspace{-0.2cm} 
	\caption{Echo signal power of the proposed RA-enabled ISAC system versus required minimum communication rate $R_{\mathrm{min}}$ for different values of $\rho$.}\vspace{-0.2cm}
	\label{fig_single_rate}
\end{figure}
First, we consider a single-user and point-target system, where the positions of the target and the user are set to $30[\sin{\frac{\pi}{2}}\cos{\frac{2\pi}{3}},\cos{\frac{\pi}{2}},\cos{\frac{\pi}{2}}\cos{\frac{2\pi}{3}}]^T$~m and $40[\sin{\frac{2\pi}{3}}\cos{\frac{\pi}{3}},\cos{\frac{2\pi}{3}},\cos{\frac{2\pi}{3}}\cos{\frac{\pi}{3}}]^T$~m, respectively.
Fig.~\ref{fig_single_rate} illustrates the echo signal power received from the target versus the required minimum communication rate $R_{\mathrm{min}}$ for different values of the antenna directivity factor $\rho$. The results show that the echo signal power decreases for all considered values of $\rho$ as the required minimum communication rate increases. This is expected since a higher communication rate requires the RAs to tilt toward the user direction in order to provide more radiation energy to support the communication link between the BS and the user, thereby inevitably reducing the directional gain in the target direction.
In addition, it can be observed that when the required minimum communication rate is relatively low, i.e., $R_{\mathrm{min}} \leq 12$~bps/Hz, a larger antenna directivity factor is more beneficial for achieving a higher echo signal power. In contrast, when the required minimum communication rate is relatively high, i.e., $R_{\mathrm{min}} \geq 16$~bps/Hz, the opposite trend is observed.
This behavior reflects the trade-off between directional gain and spatial radiation coverage.
Specifically, a larger antenna directivity factor provides a higher directional gain along the boresight direction but results in a narrower mainlobe. When the communication-rate requirement is relatively low, the RA boresights can be more closely aligned with the target direction, allowing the increased directional gain to be effectively exploited for sensing. However, as the communication-rate requirement becomes more stringent, the RA boresights have to tilt toward the communication user. If the communication and sensing directions are not well aligned, the target direction becomes increasingly off-boresight, and the sensing gain of a highly directional antenna decreases more rapidly. In contrast, a smaller directivity factor provides a wider angular radiation coverage and is therefore more robust to such directional mismatch.
Furthermore, compared with the isotropic antenna system, the RA-enabled ISAC system achieves a higher echo signal power for target detection and can satisfy more stringent communication requirements. This is because the RA-based array can flexibly tailor the spatial radiation energy distribution according to the required communication and sensing performance.

\begin{figure}[!t]  \centering
	\includegraphics[width=2.6in]{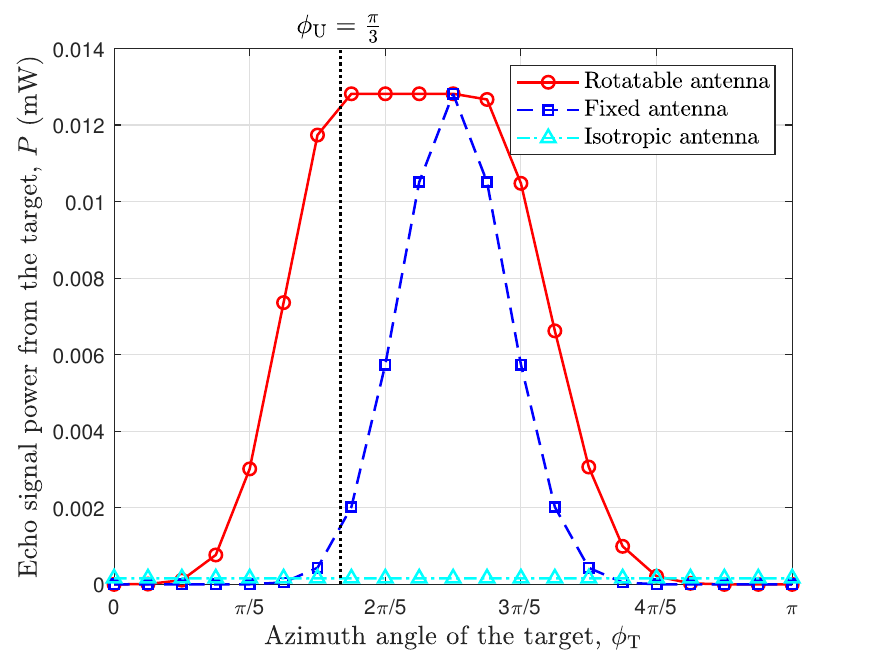}\vspace{-0.2cm} 
	\caption{Echo signal power of different ISAC systems versus target's azimuth angle  $\phi_{\mathrm{T}}$ for $R_{\mathrm{min}} = 15$ bps/Hz.}\vspace{-0.2cm}
	\label{fig_single_dis_power}
\end{figure}

\begin{figure}[!t]  \centering
	\includegraphics[width=2.6in]{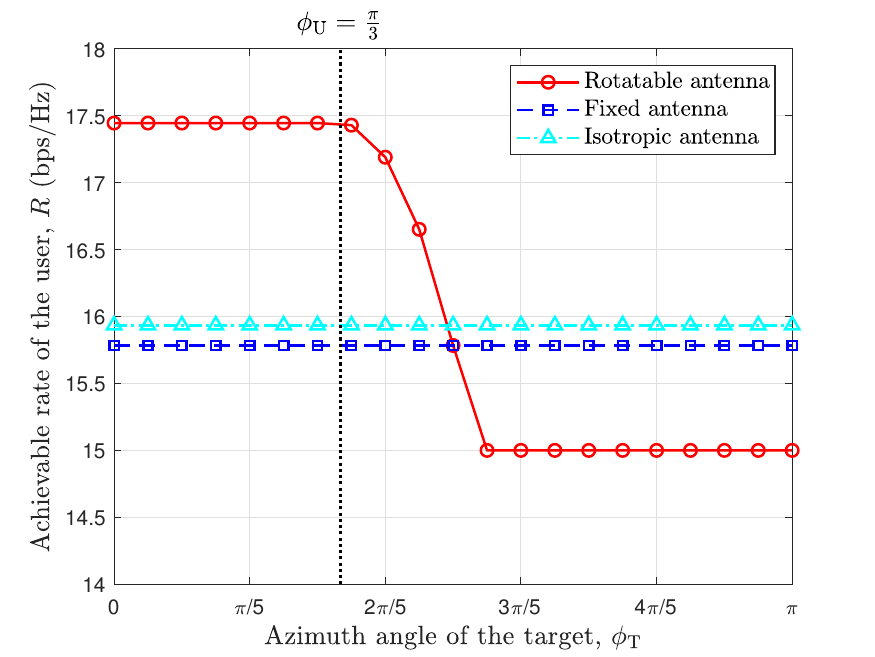}\vspace{-0.2cm} 
	\caption{Achievable rate of different ISAC systems versus target's azimuth angle  $\phi_{\mathrm{T}}$ for $R_{\mathrm{min}} = 15$ bps/Hz.}\vspace{-0.2cm}
	\label{fig_single_dis_rate}
\end{figure}
Figs.~\ref{fig_single_dis_power} and \ref{fig_single_dis_rate} present the echo signal power received from the target and the achievable rate of the user for different ISAC systems versus the target's azimuth angle $\phi_{\mathrm{T}}$ with $R_{\mathrm{min}} = 15$~bps/Hz, respectively.
As shown in Fig.~\ref{fig_single_dis_power}, the proposed RA-enabled ISAC system achieves a significantly higher echo signal power than the benchmark schemes with conventional fixed antennas and isotropic antennas for all azimuth angles, except in the two side regions of the RA array, i.e., $\phi_{\mathrm{T}}\in \left[0,\frac{\pi}{5}\right]\cup \left[\frac{4\pi}{5},\pi\right]$, where the performance advantage is diminished due to the limited antenna rotation range. Moreover, the echo signal power achieved by the proposed RA-enabled ISAC system remains approximately constant within the region $\phi_{\mathrm{T}}\in \left[\frac{\pi}{3},\frac{11\pi}{20}\right]$, which demonstrates the remarkable capability of the RA architecture to perform regional scanning and sensing. This is because, owing to the additional spatial DoFs provided by antenna rotation, the RAs can concentrate the radiated energy on a desired observation point to adaptively enhance the echo signal power, while simultaneously satisfying the communication requirement of the user.
In contrast, the fixed-antenna system has a very limited effective sensing region, since the radiated energy of the BS can only be concentrated in the fixed forward direction of the array. Although the isotropic-antenna system can provide relatively uniform sensing performance over the entire space, its radiation energy is uniformly dispersed, which results in a low echo signal power in any specific direction.

In Fig.~\ref{fig_single_dis_rate}, the achievable rate of the user always exceeds the required minimum communication rate $R_{\mathrm{min}}$. This demonstrates that even when there is a significant angular deviation between the user direction and the target direction, the RA-enabled system can still significantly improve the echo signal power while ensuring the communication quality of service of the user. This result highlights the unique benefits of the RA architecture for ISAC applications.

\subsection{Multi-User and Regional Sensing Case}
Next, we consider a general multi-user and regional sensing scenario with $K = 3$ communication users, and a horizontal sensing region located in the $x$-$z$ plane and centered at $[-10,20,30]^T$~m. Specifically, the three users are uniformly distributed in three distinct directions in front of the BS, each with a horizontal distance of 40~m, and are located on the plane $y = -10$~m. In addition, $M = M_x M_z$ spatial sampling points are uniformly selected within the sensing region, where $M_x$ and $M_z$ denote the numbers of spatial sampling points along the $x$- and $z$-axes, respectively. To ensure the same spatial sampling density for sensing regions of different sizes, we set $M_x = M_z = r+1$.
To validate the performance advantages of the proposed RA-enabled ISAC system, we consider the following three benchmark systems for comparison:
\begin{itemize}
	\item{\bf{Uniform orientation design}}: In this scheme, all RAs are constrained to have the same boresight orientation. The max-min received echo signal power of this scheme can also be obtained by Algorithm~1 by imposing the additional constraint $\vec{\mathbf{f}}_1 = \vec{\mathbf{f}}_2 = \dots = \vec{\mathbf{f}}_N$ in problem (P1). This benchmark is used to evaluate the performance loss caused by restricting all RAs to adopt a common orientation, as opposed to independently optimizing their boresight directions.
	\item{\bf{Random orientation design}}: In this scheme, the orientation of each RA is randomly generated within the rotational ranges specified in \eqref{deqn_ex2a}. The simulation results for this scheme are averaged over 100 independent realizations of the pointing vectors.
	\item{\bf{Fixed orientation design}:} In this scheme, the orientations of all RAs are fixed at their reference orientations, i.e., $\vec{\mathbf{f}}_n = \mathbf{e}_3,\; \forall n\in \mathcal{N}$.
\end{itemize}
For the above three benchmark schemes, given the RA pointing vectors, the transmit beamforming vectors for all users and the probing-signal covariance matrix are obtained by solving problem (P10).

\begin{figure}[!t]  \centering
	\includegraphics[width=2.6in]{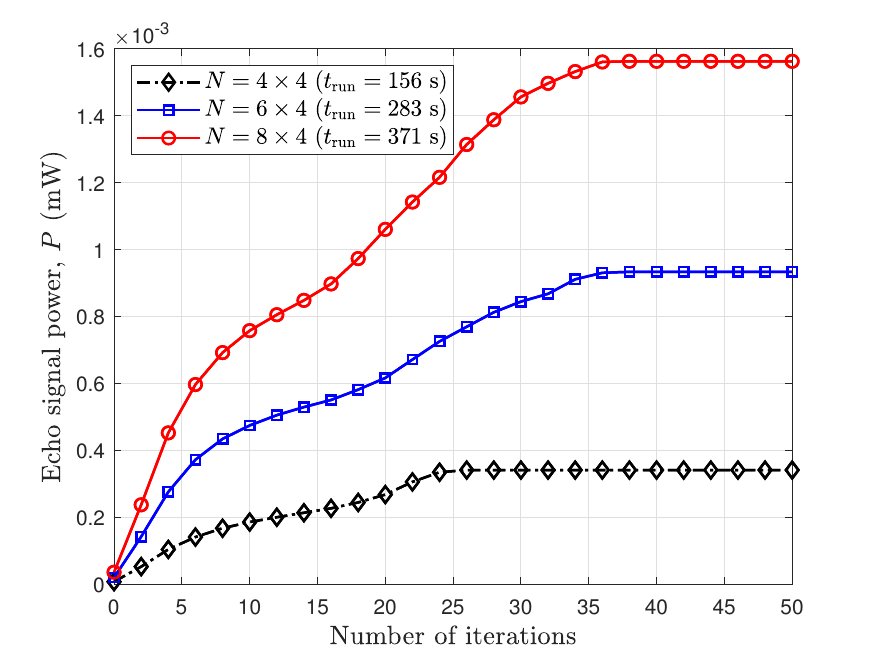}\vspace{-0.2cm} 
	\caption{Convergence behavior and actual runtime of the proposed AO algorithm for different numbers of RAs.}\vspace{-0.2cm}
	\label{fig_Convergence}
\end{figure}

First, Fig.~\ref{fig_Convergence} illustrates the convergence behavior and actual runtime of the proposed AO algorithm for different numbers of RAs. It is observed that the minimum echo signal power monotonically increases over the iterations and gradually approaches a constant value for all considered array sizes, demonstrating the stable convergence behavior of Algorithm~\ref{alg1}.
As the number of RAs increases, more iterations are generally required for convergence, since the resulting optimization problem involves more variables and higher-dimensional subproblems.
Correspondingly, the measured runtimes $t_{\mathrm{run}}$ increase from $156$~s for $N = 4\times 4$ to $283$~s and $371$~s for $N = 6\times 4$ and $N = 8\times 4$, respectively.
Moreover, increasing the number of RAs leads to a higher minimum echo signal power after convergence, as the larger array provides higher array gain and additional spatial DoFs for jointly balancing communication and sensing requirements.

\begin{figure}[!t]  \centering
	\includegraphics[width=2.6in]{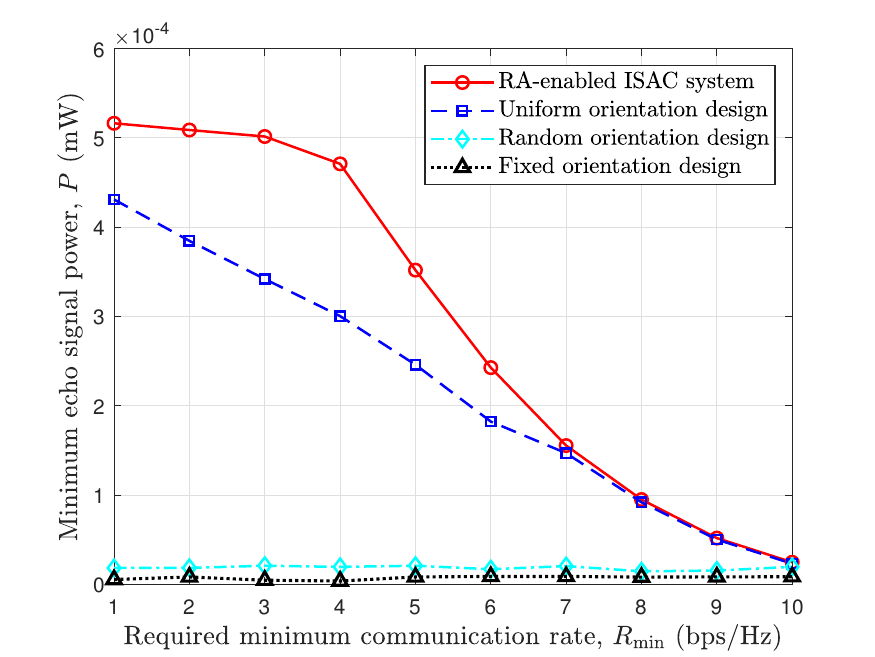}\vspace{-0.2cm} 
	\caption{Minimum echo signal power of different ISAC systems versus required minimum communication rate $R_{\mathrm{min}}$.}\vspace{-0.2cm}
	\label{fig_rate}
\end{figure}

Fig.~\ref{fig_rate} shows the minimum echo signal power achieved by the considered systems versus the required minimum communication rate $R_{\mathrm{min}}$. As can be observed, the proposed RA-enabled ISAC system consistently outperforms the benchmark systems in terms of both the minimum echo signal power over the sensing region and the supported communication rate.
This is because RAs can adaptively reconfigure the array directional gain pattern by adjusting their orientations/boresights according to the wireless environment and ISAC requirements, thus significantly enlarging the region characterizing the trade-off between communication and sensing. In contrast, although the baseline system with uniform orientation design improves the echo signal power over the random and fixed orientation designs, it cannot independently control each antenna orientation, leading to limited spatial DoFs and inferior performance compared with the proposed RA-enabled ISAC system.

\begin{figure}[!t]  \centering
	\includegraphics[width=2.6in]{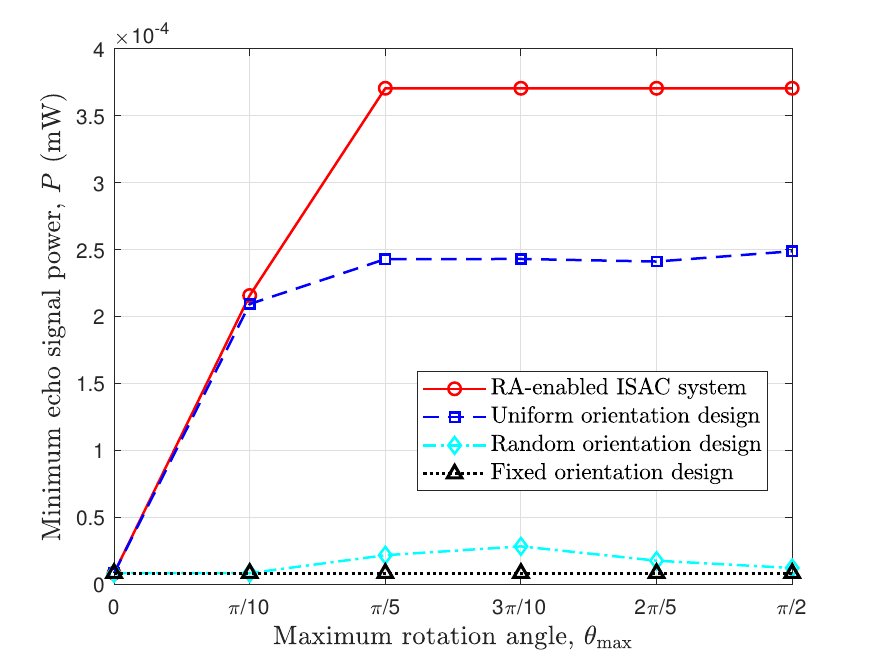}\vspace{-0.2cm} 
	\caption{Minimum echo signal power of different ISAC systems versus maximum rotation angle $\theta_{\mathrm{max}}$ for $R_{\mathrm{min}} = 5$ bps/Hz.}\vspace{-0.2cm}
	\label{fig_angle}
\end{figure}

Fig.~\ref{fig_angle} depicts the minimum echo signal power versus the maximum rotation angle $\theta_{\mathrm{max}}$. Several interesting observations can be made.
First, as $\theta_{\mathrm{max}}$ increases, the proposed RA-enabled ISAC system gains more spatial DoFs and greater flexibility to balance communication and sensing performance, thereby leading to an increase in the achieved minimum echo signal power.
Second, since the RAs with random orientations can statistically radiate power in arbitrary directions around the BS, they yield a small sensing performance gain over the fixed orientation design. However, when $\theta_{\mathrm{max}} \geq \frac{3\pi}{10}$, the echo signal power of the random orientation design decreases as $\theta_{\mathrm{max}}$ increases. This result highlights the importance and effectiveness of antenna orientation design in ISAC systems, especially when $\theta_{\mathrm{max}}$ is large, since unsuitable and uncontrolled orientation design may lead to an ISAC performance degradation.
Third, even with a small rotational range for RA orientation adjustment, the RA-enabled ISAC system can still substantially improve the echo signal power.

\begin{figure}[!t]  \centering
	\includegraphics[width=2.6in]{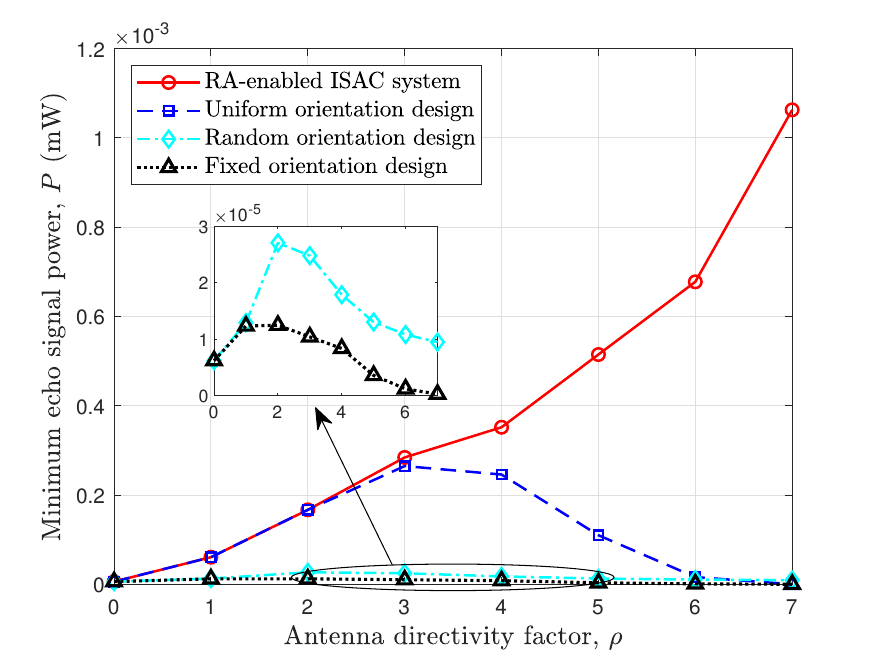}\vspace{-0.2cm} 
	\caption{Minimum echo signal power of different ISAC systems versus antenna directivity factor $\rho$ for $R_{\mathrm{min}} = 5$ bps/Hz.}\vspace{-0.2cm}
	\label{fig_direct}
\end{figure}

In Fig.~\ref{fig_direct}, we show the minimum echo signal power of the considered systems versus the antenna directivity factor~$\rho$.
As can be observed, the minimum echo signal power of the proposed RA-enabled ISAC system increases with $\rho$. This is because a larger value of $\rho$ leads to a narrower mainlobe and a higher directional gain along the antenna boresight, which enables the RA-based BS to more effectively focus its radiation power to satisfy the communication requirements of all users while sharply illuminating the sensing region of interest.
Therefore, with a larger $\rho$, the proposed RA-enabled ISAC system can generate a more adaptive and customized ISAC beam through optimized transmit beamforming and a dedicated probing signal, thereby achieving a higher echo signal power.
In contrast, the echo signal power of the uniform, random, and fixed orientation designs decreases with $\rho$ when $\rho \geq 3$, $\rho \geq 2$, and $\rho \geq 1$, respectively. Additionally, although the uniform orientation design can rotate all antennas to the same optimized orientation, its received echo signal power eventually drops to a level close to that of the fixed orientation design.
This is because, with a larger directivity factor $\rho$, the radiation power of the baseline uniform and fixed orientation design systems becomes more concentrated along the array's main pointing direction, since all antennas in both systems share the same orientation. As a result, their ability to balance communication and sensing performance is significantly weakened. For the random orientation design, a larger $\rho$ leads to a more irregular directional gain pattern and ISAC beam, making it less capable of improving ISAC performance.
The proposed RA-enabled ISAC system outperforms all baseline systems, and its performance advantage over the uniform orientation design grows as $\rho$ increases. This further demonstrates the importance of the proposed RA-enabled ISAC system for exploiting the available spatial DoFs, especially when the antennas have strong directivity.
Furthermore, for each value of $\rho$, all considered schemes employ the same radiation pattern and antenna directivity. Therefore, the performance gaps among them originate from their different capabilities to exploit the directional gain through orientation control. The increasing advantage of the proposed ISAC system with growing $\rho$ further demonstrates that element-wise orientation optimization enables the available antenna directivity to be more effectively exploited for ISAC performance enhancement.

\begin{figure}[!t]  \centering
	\includegraphics[width=2.6in]{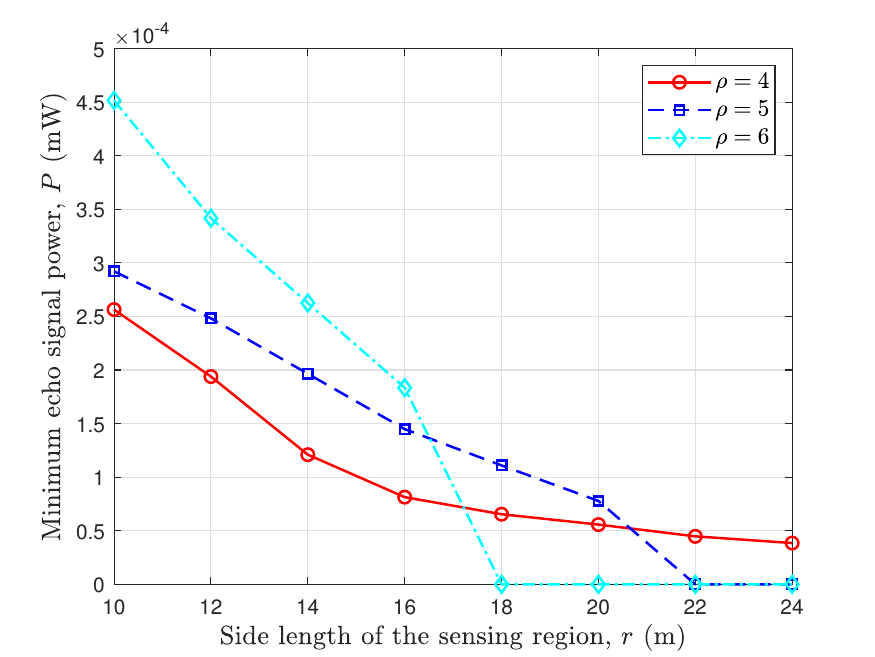}\vspace{-0.2cm} 
	\caption{Minimum echo signal power of the proposed RA-enabled ISAC system versus sensing region's side length $r$ for $R_{\mathrm{min}} = 5$ bps/Hz.}\vspace{-0.2cm}
	\label{fig_radius}
\end{figure}

In Fig.~\ref{fig_radius}, we examine the impact of the sensing region's side length $r$ on the minimum echo signal power of the proposed RA-enabled ISAC system for different values of the antenna directivity factor $\rho$.
One can observe that, as $r$ increases, the echo signal power achieved by the RA-enabled ISAC system decreases for all considered values of $\rho$. This is expected, since a larger sensing region inherently poses a significant challenge to both probing signal design and RA orientation adjustment for forming a sensing beam that covers the entire sensing region. Moreover, due to the conservation of radiation energy, enlarging the sensing region inevitably reduces the minimum echo signal power over the entire sensing region.
Furthermore, when the sensing region is small (i.e., $r\leq 16$~m), a larger $\rho$ improves the minimum echo signal power. This is because a higher directivity factor yields a narrower mainlobe and a higher directional gain along the antenna boresight, enabling the BS to concentrate more radiation power on the sensing region. However, when the sensing region becomes large (i.e., $r\geq 18$~m), increasing $\rho$ may instead reduce the minimum echo signal power. The reason for this is that a narrower beam covers a smaller spatial range, making it difficult to effectively illuminate the entire sensing region. As a result, some locations far from the RA boresight receive much less probing power, which degrades the worst-case sensing performance over the region.
This reveals a practical trade-off between radiation energy focusing and spatial coverage: higher directivity is beneficial for compact sensing regions, whereas a wider beam is preferable for larger regions to maintain more uniform sensing coverage. Therefore, the antenna directivity factor should be properly selected according to the size of the sensing region.

\begin{figure}[!t]  \centering
	\includegraphics[width=2.6in]{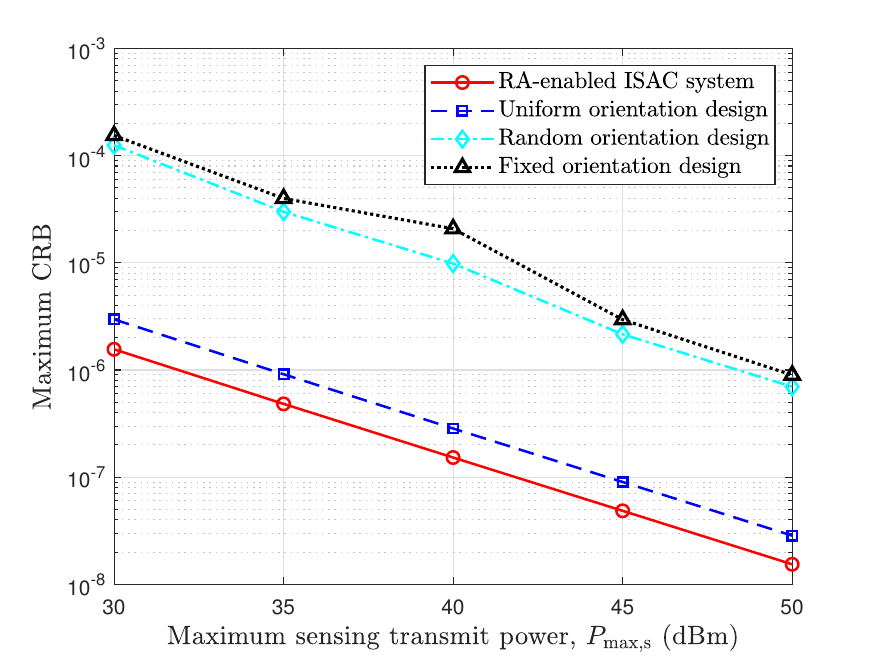}\vspace{-0.2cm} 
	\caption{Maximum CRB achieved by different ISAC systems versus maximum sensing transmit power $P_{\mathrm{max,s}}$ for $R_{\mathrm{min}} = 5$ bps/Hz.}\vspace{-0.2cm}
	\label{fig_Power_CRB}
\end{figure}

To further evaluate the sensing performance from the perspective of target estimation accuracy, Fig.~\ref{fig_Power_CRB} shows the maximum CRB versus the maximum sensing transmit power $P_{\mathrm{max,s}}$, where the CRB is derived in Appendix~\ref{CRB}.
Since the considered RA array has a relatively small aperture and the sensing region lies in the far-field of the array, the distance-dependent variations of the received signals across the array elements are relatively weak, resulting in limited Fisher information for target distance estimation. Therefore, only the CRBs associated with the two angular parameters, i.e., $\mathrm{CRB}_{\theta}$ and $\mathrm{CRB}_{\phi}$ in \eqref{deqn_ex9z}, are considered in Fig.~\ref{fig_Power_CRB}.
Specifically, the maximum CRB refers to the largest value of $\mathrm{CRB}_{\theta} + \mathrm{CRB}_{\phi}$ among all spatial sampling points within the sensing region, characterizing the worst-case direction estimation accuracy over the region.
For the CRB evaluation, the total number of sensing snapshots is set to $T = 16$. For each scheme, the RA pointing vectors and probing-signal covariance matrix are first obtained by maximizing the echo signal power through Algorithm~\ref{alg1}, based on which the CRB is subsequently evaluated over the sensing region.
It is observed that the maximum CRB decreases as $P_{\mathrm{max,s}}$ increases for all considered schemes, since a higher sensing transmit power provides stronger echo observations and thus improves the target direction estimation accuracy.
More importantly, the proposed RA-enabled ISAC design consistently achieves the lowest maximum CRB, outperforming the uniform, random, and fixed orientation designs. This result demonstrates that the proposed design, which maximizes the echo signal power, can also effectively reduce the worst-case target direction estimation uncertainty across the sensing region, thereby providing a more direct and rigorous validation of the improved sensing performance.

\section{Conclusion}\label{Conclusion}
In this paper, we investigated a new RA-enabled ISAC system to simultaneously support multi-user communication and regional sensing.
Specifically, we maximized the minimum received echo signal power over the sensing region where a potential target may be randomly located, subject to the minimum communication rate requirements of the users.
For the single-user and point-target case, we developed a tractable approximation, resulting in independent per-RA subproblems. The geometry of these subproblems was analyzed to derive closed-form solutions for the two possible cases, where the pointing vector lies either within the sector spanned by the user and target directions or on the boundary of the allowable rotation cone. For the general multi-user and regional sensing case, an efficient AO algorithm was developed to jointly optimize the transmit beamforming, the probing-signal covariance matrix, and the RA pointing vectors.
Simulation results demonstrated that the proposed RA-enabled ISAC system can significantly enlarge the region characterizing the trade-off between communication and sensing compared with various baseline schemes.
Moreover, the proposed design achieves a lower CRB for position estimation, further confirming the sensing performance gain from the perspective of target-estimation accuracy.
The results also showed that there exists a trade-off between the directional gain and the radiation coverage when choosing the antenna directivity factor.

{\appendices
	
\section{CRB Matrix for Target Position Estimation}\label{CRB}
To quantitatively characterize the target-position estimation accuracy of the considered RA-enabled ISAC system, we derive the corresponding CRB.
We parameterize the target position in spherical coordinates as $\bm{\nu}_{\mathrm{T}}\triangleq [r_{\mathrm{T}},\theta_{\mathrm{T}},\phi_{\mathrm{T}}]^T$, where $r_{\mathrm{T}} > 0$ denotes the target distance from the center of the RA array, and $\theta_{\mathrm{T}}\in [0,\pi]$ and $\phi_{\mathrm{T}}\in [0,2\pi)$ denote the polar angle and the azimuth angle, respectively. Thus, the target position in Cartesian coordinates is $\mathbf{q}_{\mathrm{T}} = r_{\mathrm{T}} \left[\sin{\theta_{\mathrm{T}}}\cos{\phi_{\mathrm{T}}},\cos{\theta_{\mathrm{T}}},\sin{\theta_{\mathrm{T}}}\sin{\phi_{\mathrm{T}}}\right]^T$.
Based on the sensing model in \eqref{deqn_ex6b} and \eqref{deqn_ex4c}, by collecting $T$ sensing snapshots and defining $\mathbf{A}(\mathbf{F},\mathbf{q}_{\mathrm{T}})\triangleq \mathbf{h}(\mathbf{F},\mathbf{q}_{\mathrm{T}}) \mathbf{h}^T(\mathbf{F},\mathbf{q}_{\mathrm{T}})$, the received echo signals can be written in matrix form as
\begin{align}
	\label{deqn_ex2z}
	\mathbf{Y}_{\mathrm{s}}\triangleq \left[\mathbf{y}_{\mathrm{s}}[1],\mathbf{y}_{\mathrm{s}}[2],\dots,\mathbf{y}_{\mathrm{s}}[T]\right] = \beta_{\mathrm{T}}\mathbf{A}\mathbf{X}_{\mathrm{s}} + \mathbf{N}_{\mathrm{s}},
\end{align}
where $\mathbf{X}_{\mathrm{s}}\triangleq \left[\mathbf{s}[1],\mathbf{s}[2],\dots,\mathbf{s}[T]\right]\in \mathbb{C}^{N\times T}$ and $\mathbf{N}_{\mathrm{s}}\triangleq \left[\mathbf{n}_{\mathrm{s}}[1],\mathbf{n}_{\mathrm{s}}[2],\dots,\mathbf{n}_{\mathrm{s}}[T]\right]\in \mathbb{C}^{N\times T}$.
For notational simplicity, $\mathbf{A}(\mathbf{F},\mathbf{q}_{\mathrm{T}})$ and $\mathbf{h}(\mathbf{F},\mathbf{q}_{\mathrm{T}})$ are denoted by $\mathbf{A}$ and $\mathbf{h}$, respectively, in the following derivation.
Consistent with the probing-signal covariance matrix defined in \eqref{deqn_ex5c}, we assume that the transmitted probing-signal matrix $\mathbf{X}_{\mathrm{s}}$ is constructed such that $\frac{1}{T}\mathbf{X}_{\mathrm{s}}\mathbf{X}_{\mathrm{s}}^H = \mathbf{S}$, which is feasible provided that $\operatorname{rank}(\mathbf{S})\le T$.
By vectorizing the received sensing signal matrix, we obtain
\begin{align}
	\label{deqn_ex3z}
	\bar{\mathbf{y}}_{\mathrm{s}} = \operatorname{vec}\hspace{-0.05cm}\left(\mathbf{Y}_{\mathrm{s}}\right) = \beta_{\mathrm{T}} \operatorname{vec}\hspace{-0.05cm}\left(\mathbf{A}\mathbf{X}_{\mathrm{s}}\right) + \operatorname{vec}\hspace{-0.05cm}\left(\mathbf{N}_{\mathrm{s}}\right).
\end{align}

Since the complex target reflection coefficient $\beta_{\mathrm{T}}$ is generally unknown, the unknown parameter vector is defined as $\bm{\xi}\triangleq \left[\bm{\nu}_{\mathrm{T}}^T,\beta_{\mathrm{T,r}},\beta_{\mathrm{T,i}}\right]^T\in \mathbb{R}^{5\times 1}$, where $\beta_{\mathrm{T,r}}\triangleq \mathrm{Re}\{\beta_{\mathrm{T}}\}$ and $\beta_{\mathrm{T,i}}\triangleq \mathrm{Im}\{\beta_{\mathrm{T}}\}$.
Since the noise follows $\operatorname{vec}\hspace{-0.05cm}\left(\mathbf{N}_{\mathrm{s}}\right)\sim \mathcal{CN}(\mathbf{0},\sigma_{\mathrm{s}}^2\mathbf{I}_{NT})$, the Fisher information matrix (FIM) for estimating the unknown parameter $\bm{\xi}$ is given by~\cite{Kay1993Fundamentals} 
\begin{align}
	\label{deqn_ex4z}
	\mathbf{J}_{\bm{\xi}} = \frac{2}{\sigma_{\mathrm{s}}^2}\mathrm{Re}\left\{\frac{\partial \bm{\mu}^H}{\partial \bm{\xi}} \frac{\partial \bm{\mu}}{\partial \bm{\xi}} \right\} = \begin{bmatrix}
	\mathbf{J}_{\bm{\nu}\bm{\nu}}& \mathbf{J}_{\bm{\nu}\bm{\beta}}\\
	\mathbf{J}_{\bm{\nu}\bm{\beta}}^T& \mathbf{J}_{\bm{\beta}\bm{\beta}}
	\end{bmatrix} \in \mathbb{R}^{5\times 5},
\end{align}
where $\bm{\mu}\triangleq \beta_{\mathrm{T}} \operatorname{vec}\hspace{-0.05cm}\left(\mathbf{A}\mathbf{X}_{\mathrm{s}}\right)$.
Here, $\mathbf{J}_{\bm{\nu}\bm{\nu}}\in \mathbb{R}^{3\times 3}$ denotes the Fisher information associated with the target-position parameters, $\mathbf{J}_{\bm{\beta}\bm{\beta}}\in \mathbb{R}^{2\times 2}$ corresponds to the unknown complex reflection coefficient, and $\mathbf{J}_{\bm{\nu}\bm{\beta}}\in \mathbb{R}^{3\times 2}$ characterizes the coupling between them.

For $i\in \{1,2,3\}$, define $\bar{\mathbf{A}}_{i}\triangleq \frac{\partial \mathbf{A}}{\partial \nu_{\mathrm{T},i}} = \bar{\mathbf{h}}_{i}\mathbf{h}^T + \mathbf{h}\bar{\mathbf{h}}_{i}^T$, where $\bar{\mathbf{h}}_{i}\triangleq \frac{\partial \mathbf{h}}{\partial \nu_{\mathrm{T},i}}$, whose explicit expression is derived in Appendix~\ref{Derivative}, and $\nu_{\mathrm{T},i}$ denotes the $i$-th entry of $\bm{\nu}_{\mathrm{T}}$ (i.e., $\nu_{\mathrm{T},1} = r_{\mathrm{T}}$, $\nu_{\mathrm{T},2} = \theta_{\mathrm{T}}$, and $\nu_{\mathrm{T},3} = \phi_{\mathrm{T}}$). 
Then, using $\mathbf{X}_{\mathrm{s}}\mathbf{X}_{\mathrm{s}}^H = T\mathbf{S}$, the entries of the FIM blocks are given by
\vspace{-0.3cm}
\begin{subequations}\label{deqn_ex5z}
\begin{align}
	\left[\mathbf{J}_{\bm{\nu}\bm{\nu}}\right]_{i,j} &= \frac{2T|\beta_{\mathrm{T}}|^2}{\sigma_{\mathrm{s}}^2} \mathrm{Re}\left\{\mathrm{Tr}\hspace{-0.05cm}\left(\bar{\mathbf{A}}_i^H \bar{\mathbf{A}}_j \mathbf{S}\right) \right\},\label{deqn_ex5z1}\\
	\left[\mathbf{J}_{\bm{\nu}\bm{\beta}}\right]_{i,:} &= \frac{2T}{\sigma_{\mathrm{s}}^2}\begin{bmatrix}
	\mathrm{Re} \left\{\beta_{\mathrm{T}}^{\ast}\mathrm{Tr}\hspace{-0.05cm} \left(\bar{\mathbf{A}}_i^H \mathbf{A} \mathbf{S}\right)\right\} \\
	-\mathrm{Im} \left\{\beta_{\mathrm{T}}^{\ast}\mathrm{Tr}\hspace{-0.05cm} \left(\bar{\mathbf{A}}_i^H \mathbf{A} \mathbf{S}\right)\right\}	
	\end{bmatrix}^T,\label{deqn_ex5z2}\\
	\mathbf{J}_{\bm{\beta}\bm{\beta}} &= \frac{2T}{\sigma_{\mathrm{s}}^2} \mathrm{Tr}\hspace{-0.05cm}\left(\mathbf{A}^H \mathbf{A} \mathbf{S}\right) \mathbf{I}_2.\label{deqn_ex5z3}
\end{align}
\end{subequations}
Since $\beta_{\mathrm{T}}$ is treated as a nuisance parameter, the equivalent FIM for estimating $\bm{\nu}_{\mathrm{T}}$ is obtained via the Schur complement as~\cite{Bekkerman2026Target}:
\begin{align}
	\label{deqn_ex6z}
	\mathbf{J}_{\bm{\nu}} = \mathbf{J}_{\bm{\nu}\bm{\nu}} - \mathbf{J}_{\bm{\nu}\bm{\beta}} \mathbf{J}_{\bm{\beta}\bm{\beta}}^{-1} \mathbf{J}_{\bm{\nu}\bm{\beta}}^T.
\end{align}
Accordingly, the CRB matrix for estimating the spherical target-position parameters is
\begin{align}
	\label{deqn_ex7z}
	\mathbf{C}_{\bm{\nu}} = \mathbf{J}_{\bm{\nu}}^{-1}.
\end{align}
By substituting \eqref{deqn_ex5z} into \eqref{deqn_ex6z}, the $(i,j)$-th entry of $\mathbf{J}_{\bm{\nu}}$, for $i,j\in \{1,2,3\}$, is derived as
\begin{align}
	\label{deqn_ex8z}
	&\left[\mathbf{J}_{\bm{\nu}}\right]_{i,j} = \left[ \left(2T|\beta_{\mathrm{T}}|^2\right)/{\sigma_{\mathrm{s}}^2} \right]\times \nonumber\\
	&\quad \; \mathrm{Re}\left\{\mathrm{Tr}\hspace{-0.05cm}\left(\bar{\mathbf{A}}_i^H \bar{\mathbf{A}}_j \mathbf{S}\right) - \frac{\mathrm{Tr}\hspace{-0.05cm} \left(\bar{\mathbf{A}}_i^H \mathbf{A} \mathbf{S}\right) \mathrm{Tr}\hspace{-0.05cm} \left(\mathbf{A}^H \bar{\mathbf{A}}_j \mathbf{S}\right)}{\mathrm{Tr}\hspace{-0.05cm}\left(\mathbf{A}^H \mathbf{A} \mathbf{S}\right)}\right\}.
\end{align}
Finally, the CRBs for estimating the target distance, polar angle, and azimuth angle are respectively given by
\begin{align}
	\label{deqn_ex9z}
	\hspace{-0.25cm}\mathrm{CRB}_{r} = \left[\mathbf{C}_{\bm{\nu}}\right]_{1,1},\; \mathrm{CRB}_{\theta} = \left[\mathbf{C}_{\bm{\nu}}\right]_{2,2},\; \mathrm{CRB}_{\phi} = \left[\mathbf{C}_{\bm{\nu}}\right]_{3,3}.\hspace{-0.2cm}
\end{align}

\section{Derivative of $\mathbf{h}(\mathbf{F},\mathbf{q}_{\mathrm{T}})$ With Respect to $\bm{\nu}_{\mathrm{T}}$}\label{Derivative}
For RA~$n$, define $d_{\mathrm{T},n} = \|\mathbf{q}_{\mathrm{T}} - \mathbf{w}_n\|$, $\vec{\mathbf{q}}_{{\mathrm{T}},n} = \frac{\mathbf{q}_{{\mathrm{T}}} - \mathbf{w}_n}{d_{\mathrm{T},n}}$, and $c_{\mathrm{T},n} = \vec{\mathbf{f}}_n^T \vec{\mathbf{q}}_{{\mathrm{T}},n}$.
According to \eqref{deqn_ex2b} and \eqref{deqn_ex3b}, for $c_{\mathrm{T},n} > 0$, the channel coefficient between RA~$n$ and the target can be written as
\begin{align}
	\label{deqn_ex1x}
	h_n = \frac{\sqrt{\beta_0 G_{\max}}}{d_{\mathrm{T},n}} c_{\mathrm{T},n}^{\rho} \exp\left(-j\frac{2\pi}{\lambda} d_{\mathrm{T},n}\right).
\end{align}
Then, for $i\in \{1,2,3\}$, taking the logarithmic derivative of $h_n$ with respect to $\nu_{\mathrm{T},i}$, we obtain
\begin{align}
	\label{deqn_ex2x}
	\frac{1}{h_n}\frac{\partial h_n}{\partial \nu_{\mathrm{T},i}}  = -\left(\frac{1}{d_{\mathrm{T},n}} + j\frac{2\pi}{\lambda}\right)\frac{\partial d_{\mathrm{T},n}}{\partial \nu_{\mathrm{T},i}} + \frac{\rho}{c_{\mathrm{T},n}} \frac{\partial c_{\mathrm{T},n}}{\partial \nu_{\mathrm{T},i}},
\end{align}
where
\begin{align}
	\frac{\partial d_{\mathrm{T},n}}{\partial \nu_{\mathrm{T},i}} &= \vec{\mathbf{q}}_{{\mathrm{T}},n}^T \frac{\partial \mathbf{q}_{\mathrm{T}}}{\partial \nu_{\mathrm{T},i}}, \label{deqn_ex3x}\\
	\frac{\partial c_{\mathrm{T},n}}{\partial \nu_{\mathrm{T},i}} &= \frac{1}{d_{\mathrm{T},n}} \vec{\mathbf{f}}_n^T\left(\mathbf{I} - \vec{\mathbf{q}}_{{\mathrm{T}},n}\vec{\mathbf{q}}_{{\mathrm{T}},n}^T\right) \frac{\partial \mathbf{q}_{\mathrm{T}}}{\partial \nu_{\mathrm{T},i}}, \label{deqn_ex4x}
\end{align}
denote the derivatives of the distance $d_{\mathrm{T},n}$ and the directional projection $c_{\mathrm{T},n}$ with respect to $\nu_{\mathrm{T},i}$, respectively. In addition, $\frac{\partial \mathbf{q}_{\mathrm{T}}}{\partial \nu_{\mathrm{T},i}}$ denotes the derivative of $\mathbf{q}_{\mathrm{T}}$ with respect to $\nu_{\mathrm{T},i}$.

Next, substituting \eqref{deqn_ex3x} and \eqref{deqn_ex4x} into \eqref{deqn_ex2x}, the channel gradient can be expressed as
\begin{align}
	\label{deqn_ex5x}
	\hspace{-0.25cm}\frac{\partial h_n}{\partial \nu_{\mathrm{T},i}} &\hspace{-0.05cm}=\hspace{-0.05cm} h_n\hspace{-0.05cm} \left[-\hspace{-0.05cm}\left(\hspace{-0.05cm}\frac{1}{d_{\mathrm{T},n}} \hspace{-0.05cm}+\hspace{-0.05cm} j\frac{2\pi}{\lambda}\right)\hspace{-0.05cm}\vec{\mathbf{q}}_{{\mathrm{T}},n}^T \hspace{-0.05cm}+\hspace{-0.05cm} \frac{\rho \vec{\mathbf{f}}_n^T\hspace{-0.1cm} \left(\mathbf{I} \hspace{-0.05cm}-\hspace{-0.05cm} \vec{\mathbf{q}}_{{\mathrm{T}},n}\vec{\mathbf{q}}_{{\mathrm{T}},n}^T\right)}{d_{\mathrm{T},n} c_{\mathrm{T},n}}\right] \hspace{-0.1cm}\frac{\partial \mathbf{q}_{\mathrm{T}}}{\partial \nu_{\mathrm{T},i}}\nonumber\\
	&\hspace{-0.05cm}=\hspace{-0.05cm} h_n\hspace{-0.05cm}\left[\frac{\rho}{d_{\mathrm{T},n} c_{\mathrm{T},n}}\vec{\mathbf{f}}_n^T \hspace{-0.05cm}-\hspace{-0.05cm} \left(\frac{\rho + 1}{d_{\mathrm{T},n}} + j\frac{2\pi}{\lambda} \right)\hspace{-0.05cm}\vec{\mathbf{q}}_{{\mathrm{T}},n}^T\right]\hspace{-0.1cm} \frac{\partial \mathbf{q}_{\mathrm{T}}}{\partial \nu_{\mathrm{T},i}}.\hspace{-0.05cm}
\end{align}
Therefore, for $i\in \{1,2,3\}$, we have
\begin{align}
	\label{deqn_ex6x}
	\bar{\mathbf{h}}_{i} = \frac{\partial \mathbf{h}}{\partial \nu_{\mathrm{T},i}} = \left[ \frac{\partial h_1}{\partial \nu_{\mathrm{T},i}}, \frac{\partial h_2}{\partial \nu_{\mathrm{T},i}}, \dots, \frac{\partial h_N}{\partial \nu_{\mathrm{T},i}} \right]^T\hspace{-0.05cm}.
\end{align}

}

\ifCLASSOPTIONcaptionsoff
  \newpage
\fi

\bibliographystyle{IEEEtran}
\bibliography{RA_ISAC}

\end{document}